\documentclass[prb,aps,floatfix,superscriptaddress]{revtex4-2}

\usepackage{float}
\usepackage{color}
\usepackage{amsmath}
\usepackage{latexsym}
\usepackage{amssymb}
\usepackage{amsmath}
\usepackage{amsfonts}
\usepackage{mathrsfs}
\usepackage{graphics,epstopdf}
\usepackage{placeins}
\usepackage[colorlinks=true, citecolor=blue, urlcolor=blue, linkcolor=blue ]{hyperref}
\usepackage{epsf,graphics,graphicx}

\date{\today}

\usepackage{physics}
\usepackage{bigints}
\usepackage{multirow}
\usepackage{braket}
\usepackage{makecell}

\def\derivative#1#2{\frac{\partial #1}{\partial #2}}

\def\derivativec#1#2#3{\left(\frac{\partial #1}{\partial #2}\right)_{#3}}

\def\kgam{k_{\gamma}\gamma}
\def\K{\ \text{K}}

\begin{document}

\title{Multicomponent Grain Boundary Segregation Dilute-Limit Model and Its Effect on Nanocrystalline Stability}

\author{Georgiy Marchiy}
\email{georgiy.marchiy@mail.ioffe.ru}
\author{Feodor Kuznetsov}
\author{Dmitry Samsonov}
\author{Eugene Mukhin}
\affiliation
{Ioffe Institute, Saint-Petersburg, Russia}
\date{\today} 

\begin{abstract}

Grain boundary segregation of solutes is a powerful tool for stabilizing nanocrystalline materials against grain growth. However, analytical models have been largely restricted to binary systems, leaving the stabilization potential of multicomponent alloys poorly understood. In this work, we develop an analytical framework for an arbitrary number of solutes in the dilute limit of grain boundary concentrations based on segregation energy spectra computed via molecular dynamics.  We demonstrate that multicomponent segregation synergistically expands the range of nanocrystalline alloys that are thermodynamically stable against both grain growth and phase separation. Crucially, the model reveals that the cosegregation of almost any solutes enhances stability against phase separation, while the cosegregation of solutes with anti-correlated segregation spectra can suppress temperature-driven desegregation.  This framework proposes a predictive tool for the rational design of thermally stable nanocrystalline alloys using multicomponent doping. 
\end{abstract}

\maketitle

\section{Introduction}
Nanocrystalline materials attract significant interest due to their enhanced strength caused by the Hall-Petch effect, as well as a high volume fraction of grain boundaries that significantly influence bulk material properties~\cite{gleiter}. Interest is also driven by the fact that many thin-film metal coatings possess a nanocrystalline structure~\cite{petrov2003}. However, all nanocrystalline materials exhibit an inherent instability driven by grain growth~\cite{peng_thermal_2017}, which severely limits their utility in high-temperature applications~\cite{chookajorn_design_2012}. In materials with low melting points, grain growth can occur even at room temperature.
\par
A specific example of such detrimental thermal effect is silver reflective coatings. These coatings are highly valuable owing to the exceptional reflectivity of silver over a broad spectral range. Nevertheless, their application in fields such as solar energy~\cite{antonaia_adhesion_2014, gledhill_hipims_2019} (including solar concentrators and internal reflective surfaces of solar cells), space optics~\cite{risse_novel_2008, sheikh_durable_2008}, industrial high-temperature reflectors~\cite{gledhill_hipims_2019}, and optical plasma diagnostics in thermonuclear reactors~\cite{samsonov_large-scale_2022} (such as Thomson scattering diagnostics in the ITER tokamak) is heavily constrained by post-annealing degradation. Thermally induced grain growth leads to increased surface roughness, the formation of cavities, up to the complete agglomeration of the thin films~\cite{jacquet_grain_2016}.  Furthermore, it induces structural stresses~\cite{chason_tutorial_2016} that risk delamination of both the silver layer itself and its protective anticorrosive coatings~\cite{tereshenko2025}.
\par
Grain growth is driven by the system's tendency to minimize its internal energy by reducing the fraction of grain boundaries (GB)~\cite{peng_thermal_2017}, which possess excess energy. However, as demonstrated by Weismueller~\cite{weissmuller_alloy_1993}, the energy of GBs can be reduced by introducing a solute that segregates at the GBs. The preferential segregation of such a solute to the GBs decreases the system’s energy, thereby potentially making the existence of a certain GB fraction energetically favorable. Later, Trelewicz and Schuh~\cite{trelewicz_grain_2009} extended Weismueller's model to high solute concentrations within the framework of the regular solution approximation. Recent advances in this field are associated with the use of atomistic simulation methods for polycrystals and the consideration of the heterogeneous structure of grain boundaries, described by a distribution (spectrum) of segregation energies~\cite{matson_overview_2026}. However, all these models are strictly limited to binary mixtures, whereas many technologically important alloys are multicomponent. As noted by Wagih et al.~\cite{wagih_designing_2025}, the segregation of multiple solutes can open new pathways for the design of nanocrystalline materials: the spectral models predict the possibility of both competitive segregation -- where solute atoms compete for the same sites in GBs, displacing one another -- and cooperative segregation, where solute atoms occupy different sites within GBs, thereby enhancing the thermal stabilization effect. Nevertheless, to the best of our knowledge, no thermodynamic model currently exists that describes the actual effect of multicomponent segregation on nanocrystalline stability. In particular, we have found no published analytical spectral model describing the segregation of multiple solutes at GBs; while Wagih et al.~\cite{wagih_designing_2025} employed more computationally expensive Monte Carlo simulation methods to calculate segregation isotherms in a multicomponent model. This work aims to address this gap by developing an analytical spectral framework for multicomponent segregation.
\par
Guttmann~\cite{guttmann_1975, guttmann_1995} pioneered the description of multi-solute GB segregation, suggesting the possibility of both competitive and non-competitive interactions. However, his analysis does not explicitly account for the spectral distribution of GB energies. Guttmann analyzes the impact of competition among solutes for the same sites, as well as effects related to attraction and repulsion between solutes of different types, using the regular solution approximation. Nevertheless, as noted by Guttmann~\cite{guttmann_1995} and later detailed by Wagih et al.~\cite{wagih_designing_2025}, solutes can significantly influence each other's segregation even in the absence of solute--solute interactions, solely due to competition for GB sites. Therefore, at this stage, our analysis is confined to determining the impact of spectral dependence alone on multicomponent solute segregation and the corresponding thermal stabilization effect. 
\par
In this work, we present the derivation of an analytical model for multicomponent segregation of substitutional solutes without considering solute--solute interactions and perform theoretical and numerical analyses of the impact of multicomponent segregation on nanocrystalline stability. The proposed model serves as a baseline framework that excludes solute–solute interactions; however, it still yields accurate results in the dilute limit of GB solute concentration (except for systems characterized by strong attractive solute--solute interactions, as demonstrated in previous work for a binary system~\cite{marchiy_spectral_2025}).
\par
The paper is organized as follows. Section~\ref{sec:model} presents the theoretical framework, beginning with the derivation of the Gibbs free energy for multicomponent segregation of substitutional solutes in the dilute limit (Sec.~\ref{subsec:gibbs}), followed by the formulation of the segregation isotherm (Sec.~\ref{subsec:isotherm}) and the equation for nanocrystalline stability (Sec.~\ref{subsec:stability_eq}). This framework is then extended by the formulation of metastability (Sec.~\ref{subsec:phi_M_0K}) and absolute stability criteria (Sec.~\ref{subsec:abs_stability_criterion}), including a discussion of the kinetic limitations of second-phase nucleation (Sec.~\ref{subsec:nucleation}). In Section~\ref{sec:identical_coseg} we analyze a simplified symmetric model to isolate the collective effects of cosegregation. It is shown that solutes with uncorrelated spectra provide the highest potential for enhancing stability with respect to phase separation, whereas solutes with anti-correlated spectra may be highly effective in suppressing temperature-driven solute desegregation and subsequent grain growth. Section~\ref{sec:methods} details the atomistic simulation methods used to parameterize the model for real materials, including molecular dynamics and machine-learning-accelerated calculations of segregation spectra. In Section~\ref{sec:results} we address the details of the numerical solution of the model's equations and apply the model to nanocrystalline silver as a case study, demonstrating that cosegregation can expand the range of thermodynamically stable alloys. A key finding is that the multi-solute cosegregation reduces the driving force for phase separation by lowering the system's free energy, unless highly stable intermetallic compounds are formed. These insights provide actionable guidelines for using multicomponent segregation to engineer stable, precipitation-resistant nanocrystalline structures.
\section{Cosegregation Model}\label{sec:model}

Formulating the system’s Gibbs free energy is essential to understanding solute segregation and confirming nanocrystalline stability. This study derives the Gibbs free energy for the spectral model for multicomponent segregation of substitutional solutes, then determines the equilibrium state, expressed as the spectral segregation isotherm, by minimizing the energy functional with respect to solute distribution. With this formulation one can determine the grain size corresponding to minimum system Gibbs free energy. Finally, comparing this minimum Gibbs energy with the energies of competing phases (solid solution, solute--rich and intermetallic precipitates) allows us to evaluate the stability of the nanocrystalline state against decomposition~\cite{hildebrandt_predicting_1994}.
\par
The physical layout of the model assumes a polycrystal divided into grain interiors and grain boundary (GB)~\cite{matson_overview_2026}. The grain matrix is homogeneous and contains equivalent sites, whereas the GB comprises a network of randomly distributed sites with distinct local atomic environments.   Sites and atomic species are indexed by Latin subscripts i and Greek superscripts $\alpha$, respectively. The grain interior is denoted by the subscript $c$. Accordingly, the segregation energy $E^{\alpha}_i$ is defined as the change in potential energy when a single solute atom $\alpha$ is transferred from the grain interior to a specific GB site $i$.
\subsection{Gibbs Free Energy of Segregation}\label{subsec:gibbs}
Within this model the Gibbs free energy for segregation of $m$ solutes, omitting solute--solute interactions, is expressed as (see Appendix~\ref{app:gibbs}):

\begin{equation}\label{eq:gibbs_m}
G_{seg} = \sum_i F_i\sum_{\alpha}\left[ X^{\alpha}_i E^{\alpha}_i +kT X^{\alpha}_i\ln (X^{\alpha}_i)\right]\,,
\end{equation}

where  $F_i = f_{gb}\rho_i + (1-f_{gb})\delta_{i,c}$ is the fraction of $i$ sites in the polycrystal, $f_{gb}$ is the fraction of GB atoms, $\rho_i$ is the fraction of $i$ sites in GB, and $X^{\alpha}_i$ is the segregation concentration of solute $\alpha$ at $i$ sites. In this expression exclusively, the superscript $\alpha$ includes both solute and matrix $A$ atoms, with $E^A_i = 0$ and $E_c^{\alpha} = 0$ established by definition.
\par
The Gibbs free energy of mixing for the nanocrystalline phase is then expressed as the sum of the solute energy in the grain interior, the segregation energy $G_{seg}$, and the structural GB energy per atom $\kgam$:

\begin{equation}\label{eq:gibbs_mix}
G_{nc} = \sum_{\alpha} X_{tot}^{\alpha}E_{sol}^{\alpha/A} + G_{seg} + f_{gb}\kgam\,,
\end{equation}

where $E_{sol}^{\alpha/A}$ represents the energy difference for a solute atom transitioning from its elemental lattice state to the grain interior. 
\subsection{Segregation Isotherm}\label{subsec:isotherm}
Once the Gibbs free energy is defined, the solute atomic configurations that minimize this energy for a fixed grain size can be determined. Throughout this minimization, the total concentration of solutes ($ X^{\alpha}_{tot}$) is kept constant, acting as an external parameter of the system:

\begin{equation}
\sum_i F_i X^{\alpha}_i = X^{\alpha}_{tot}\,.
\end{equation}

Thus, determining the solute distribution within a polycrystal reduces to a constrained minimization problem. To find this minimum, we employ the method of Lagrange multipliers, following our previous approach~\cite{marchiy_spectral_2025}. The resulting Lagrange function is given by:

\begin{equation}\label{eq:lagrange}
L = G_{seg} - \sum_{\alpha}\mu^{\alpha}\sum_i F_i X^{\alpha}_i\,.
\end{equation}

Setting the first derivative to zero yields the necessary equilibrium condition: 

\begin{equation}\label{eq:mu}
E^{\alpha}_i + kT\ln \frac{X^{\alpha}_i}{1-\sum_{\beta}X^{\beta}_i} - \mu^{\alpha} = 0\,.
\end{equation}

Expressing the system in terms of the following variables $\varepsilon^{\alpha}_i=\exp\left(-\frac{E^{\alpha}_i}{kT}\right)$ and $M^{\alpha}=\exp\left(\frac{\mu^{\alpha}}{kT}\right)$ leads to the system of equations in the form: 

\begin{equation}
 X^{\alpha}_i = \frac{\varepsilon^{\alpha}_i M^{\alpha}}{1+\sum_{\beta}\varepsilon^{\beta}_i M^{\beta}}
\end{equation}

 or via energy:

\begin{equation}\label{eq:X}
 X^{\alpha}_i = \frac{\exp\left(-\frac{E^{\alpha}_i-\mu^{\alpha}}{kT}\right)}{1+\sum_{\beta}\exp\left(-\frac{E^{\beta}_i-\mu^{\beta}}{kT}\right)}\,.
\end{equation}

Note that Eq.~\eqref{eq:X} can be recast into the traditional form of the Guttmann segregation isotherm (omitting solute--solute interactions)~\cite{guttmann_1995}:

\begin{equation}\label{eq:X_Guttman}
 \frac{X^{\alpha}_i}{1-\sum_{\beta}X^{\beta}_i} = \frac{X^{\alpha}_c}{1-\sum_{\beta}X^{\beta}_i}\exp \left(-\frac{E^{\alpha}_i}{kT}\right)\,.
\end{equation}

This establishes the equilibrium solute distribution configuration for the polycrystal. As indicated by Eq.~\eqref{eq:X}, site competition occurs when multiple solutes favor segregation at the same grain boundary sites, reducing each solute's local concentration relative to its uncompetitive, individual segregation. Conversely, competitive effects are minimized if a site exhibits a negative segregation energy for only one species while remaining positive for the rest.
\subsection{Nanocrystalline Stability}\label{subsec:stability_eq}
When segregation is energetically favorable enough for the associated Gibbs energy to compensate for the excess GB structural energy, an equilibrium state emerges featuring solute--decorated GBs~\cite{trelewicz_grain_2009}. This equilibrium configuration is determined by minimizing the total Gibbs free energy with respect to the GB fraction $f_{gb}$:

\begin{equation}
\pdv{G_{nc}}{f_{gb}} = 0\,.
\end{equation}

This equation yields the following relation (derivation is provided in Appendix~\ref{app:eq1}):

\begin{equation}\label{eq:eq1}
\sum_i\rho_i \ln(1-\sum_{\alpha}X^{\alpha}_i) - \ln(1-\sum_{\alpha}X^{\alpha}_i) + \frac{\kgam}{kT} = 0\,.
\end{equation}

Hereafter, we refer to Eq.~\eqref{eq:eq1} as the Stability Equation. For convenience, we rewrite it as a function of $M$ and $\varepsilon$, denoting the left side of Eq.~\eqref{eq:eq1} as $I_{T}$, which represents a function of $\mu^{\alpha}$ at a given temperature $T$:

\begin{equation}\label{eq:eq1_M}
I_T(\mu^{\alpha}) = \sum_i\rho_i \ln(1+\sum_{\alpha}M^{\alpha}\varepsilon^{\alpha}_i) - \ln(1+\sum_{\alpha}M^{\alpha}) - \frac{\kgam}{kT} = 0\,.
\end{equation}

This relation imposes a condition of the type $\mu^{\beta} = f(\{\mu^{\alpha}\}_{\alpha\neq \beta})$, allowing one chemical potential to be eliminated from a system of equations.
\par
Solving the problem requires additional $m-1$ equations, which can be derived from the relations describing solute distribution between grain interior and GBs:

\begin{equation}\label{eq:X_balance}
f_{gb}X^{\alpha}_{gb} + (1-f_{gb})X^{\alpha}_{c} = X^{\alpha}_{tot}\,,
\end{equation}

where $X^{\alpha}_{gb} = \sum_i \rho_i X_i^{\alpha}$ is the mean GB $\alpha$ solute concentration,

\begin{equation}\label{eq:fgb}
f_{gb} = \frac{X^{\alpha}_{tot}-X^{\alpha}_{c}}{X^{\alpha}_{gb}-X^{\alpha}_{c}}\,.
\end{equation}

To close the system, the remaining  $m-1$ necessary equations can be chosen from the following set of $m(m-1)/2$ equations:

\begin{equation}\label{eq:eq2}
 \frac{X^{\beta}_{tot}-X^{\beta}_{c}}{X^{\beta}_{gb}-X^{\beta}_{c}}= \frac{X^{\alpha}_{tot}-X^{\alpha}_{c}}{X^{\alpha}_{gb}-X^{\alpha}_{c}},\qquad \alpha \neq \beta
\end{equation}

 or their independent linear combinations. Hereafter, we refer to Eqs.~\eqref{eq:eq2} as the Mass Balance Equations, which capture a key physical constraint: in multicomponent systems, a polycrystal's grain size and, therefore, the corresponding GB segregation capacity cannot adjust to satisfy the individual optimum concentration of every solute simultaneously. Rather, Eq.~\eqref{eq:eq1} and Eqs.~\eqref{eq:eq2} force an energetic compromise among all coexisting solutes. Because of this thermodynamic interdependence, adding a new solute to an already balanced system alters the GB concentrations of existing elements, prompting a redistribution controlled by the concentration of the newly introduced species.
 \par
The existence of roots for Eq.~\eqref{eq:eq1} indicates that the system reaches a local minimum of the Gibbs free energy in the nanocrystalline state. However, phase separation may be energetically more favorable, in which case the nanocrystalline state is considered metastable. If, even after precipitation, the solute remaining at the grain boundaries is sufficient to suppress grain growth, the nanocrystalline state is defined as absolutely stable.
\subsection{Metastability Criterion}\label{subsec:phi_M_0K}
In the ground state solutes occupy sites corresponding to the most negative segregation energies. Each site is preferentially occupied by specific solute species that minimizes the energy, allowing each site to be characterized by the energy $\tilde{E}_i = \min_{\alpha}(E^{\alpha}_i)$. The lowest possible GB energy at 0 K will be equal to the sum of the GB structural energy per atom, $\kgam$, and all negative $\tilde{E}_i$ values per atom: $ \kgam +\sum_i^{\tilde{E}_i<0} \rho_i\tilde{E}_i$. If this value is negative, the solute can potentially suppress grain growth. Therefore, the criterion for metastability at 0 K requires the stability score $\Phi^{M}_{0K}$ to exceeds unity:

\begin{equation}\label{eq:phi_m0}
 \Phi^{M}_{0K}(\{\alpha\}) =- \frac{1}{\kgam}\sum_i^{\tilde{E}_i<0} \rho_i\tilde{E}_i>1\,.
\end{equation}

Equation.~\eqref{eq:phi_m0} resembles the stability criterion obtained by Wagih and Schuh for binary systems~\cite{wagih_thermodynamics_2021}:

\begin{equation}
\Phi^M_{0K}(\alpha)=-\frac{1}{k_\gamma \gamma} \sum_{i}^{E_i^{\alpha}<0} \rho_iE_i^{\alpha}>1\,.
\end{equation}
Notably, the relation $\Phi^{M}_{0K}(\{\alpha\}) \ge \max_{\alpha}\{\Phi^{M}_{0K}(\alpha)\}$ demonstrates that the cosegregation of multiple solutes can potentially enhance thermodynamic stability. The maximum increase in $\Phi^{M}_{0K}$ occurs when the respective segregation spectra are anti-correlated, whereas the stability score remains unchanged when the spectra are fully correlated.

\subsection{Absolute Stability Criterion}\label{subsec:abs_stability_criterion}
As noted above, the segregated state can exhibit metastability.  While the solute segregated within the GBs stabilizes the nanocrystalline structure, the system can remain unstable with respect to phase separation. Precipitation of the segregated solute depletes the GB solute concentration, potentially disrupting nanocrystalline stability. Assessing absolute stability requires comparing the Gibbs free energy of mixing of the nanocrystalline phase ($G_{nc}$) against all alternative phases, where absolute thermodynamic stability is achieved for phases residing on the convex hull of the $G_{mix}(X^{\alpha}_{tot})$ diagram~\cite{hildebrandt_predicting_1994}.
\par
It should be noted that the Lagrange multipliers introduced in Eq.~\eqref{eq:lagrange} equal the solute chemical potentials in the nanocrystalline phase ($\mu^{\alpha}_{nc}$) up to a constant offset. Indeed, chemical potentials are defined as the derivatives of the Gibbs free energy with respect to concentration:

\begin{equation}
\mu_{nc}^{\alpha}=\frac{dG_{nc}}{dX^{\alpha}_{tot}} = E_{sol}^{\alpha/A}  + \frac{dG_{seg}}{dX^{\alpha}_{tot}}\,.
\end{equation}

Differentiating Eq.~\eqref{eq:lagrange} with respect to $X^{\alpha}_i$ yields:

\begin{equation}
\pdv{G_{seg}}{X^{\alpha}_i}=F_i \mu^{\alpha}\,,
\end{equation}

thus,

\begin{equation}
\frac{dG_{seg}}{dX^{\alpha}_{tot}}  = \sum_i \sum_{\beta}\pdv{G_{seg}}{X_i^{\beta}}\frac{dX_i^{\beta}}{dX^{\alpha}_{tot}} = \sum_i \sum_{\beta}F_i \mu^{\beta}\frac{dX_i^{\beta}}{dX^{\alpha}_{tot}} = \sum_{\beta} \mu^{\beta}\frac{dX_{tot}^{\beta}}{dX^{\alpha}_{tot}} = \mu^{\alpha}\,.
\end{equation}
Finally we obtain $\mu_{nc}^{\alpha} = E_{sol}^{\alpha/A}  + \mu^{\alpha}$. Therefore, for $X^{\alpha}_{tot} > X^{\alpha}_c$, the Gibbs energy of the nanocrystalline phase can be written as $G_{nc} = G_0 + \sum_{\alpha}\left(E_{sol}^{\alpha/A} + \mu^{\alpha}\right)X^{\alpha}_{tot}$, where $G_0$ is obtained by subtracting the derived expression from Gibbs energy (Eq.~\eqref{eq:gibbs_mix}) using Eqs.~\eqref{eq:mu} and~\eqref{eq:eq1}: $G_0 = kT\ln(1 - \sum_{\alpha}X_c^{\alpha}) = -kT\ln(1 + \sum_{\alpha} M^{\alpha})$.
\par
It should be noted that for positive $\mu^{\alpha}_{nc}$, the minimum energy is achieved at a point corresponding to $f_{gb}=0$. Consequently, under $\mu^{\alpha}_{nc}>0$ conditions, the nanocrystalline state is necessarily metastable. In the absence of intermetallic compounds, and provided the enthalpy of mixing is not strongly negative (typical for non-intermetallic systems), negative chemical potentials directly indicate thermodynamic stability. 
\par
When systems form intermetallic compounds, thermodynamic analysis becomes more intricate. Consider the intermetallic compound $A_{1-x^{B}_l}B_{x^{B}_l}$  with the lowest solute stoichiometry $x_l^B$ lying on the convex hull possessing the lowest chemical potential among all competing phases. It should be noted that the region of low solute concentrations is of primary interest. Therefore, to verify absolute stability, the chemical potentials of the solutes in the nanocrystalline phase must be lower than those in the corresponding intermetallic compound: $\mu_{nc}^{B} < E_{form}(A_{1-x^{B}_l}B_{x^{B}_l})/x^{B}_l$, where $E_{form}$ is the formation energy of this intermetallic compound. 
\par
Beyond these matrix-solute binaries, complex intermetallics consisting of multiple solute elements can also exist. In ternary systems, a  $B_{1-z}C_z$ intermetallic compound will not precipitate within the low-concentration regime if the following condition is met:

\begin{equation}\label{eq:BC_inter_cond}
E_{form}(B_{1-z}C_z) > \frac{E_{form}(A_{1-x^B_l}B_{x^B_l})}{x^B_l}(1-z) + \frac{E_{form}(A_{1-x^C_l}C_{x^C_l})}{x^C_l}z\,.
\end{equation}

Generalization to multicomponent systems yields:

\begin{equation}\label{eq:multi_inter_cond}
E_{form}(A_{x_0}B^1_{x_1}\dots B^m_{x_m}) > \sum_{k=1}^{m}\frac{E_{form}(A^{\,}_{1-x^{B^k}_l}B^{k}_{x^{B^k}_l})}{x^{B^k}_l}x_k\,,
\end{equation}

where $x_k \in [0, 1)$ and $\sum_{k=0}^{m}x_k = 1$. If the condition in Eq.~\eqref{eq:multi_inter_cond} is satisfied, the stability can be evaluated via the criterion  $\mu_{nc}^{\alpha} < E_{form}(A_{1-x^{\alpha}_m}\alpha_{x^{\alpha}_m})/x^{\alpha}_m$ for all $\alpha$.  Otherwise, the analysis requires constructing a phase diagram. We further focus on the cases where Eq.~\eqref{eq:BC_inter_cond} is met. Let us define the reference solute energy (the energy of a solute atom in its most favorable phase) as $E_{ref}^{\alpha} = E_{sol}^{\alpha/A} - E_{form}(A_{1-x^{\alpha}_m}\alpha_{x^{\alpha}_m})/x^{\alpha}_m$. Then, the necessary condition for absolute stability, $\mu^{\alpha} < -E^{\alpha}_{ref}$, must be satisfied for all $\alpha$. It should be noted that in the present model of the nanocrystalline phase, the chemical potential is independent of composition; therefore, the absolute stability condition becomes a function of temperature alone.
\par 
This condition is easily verified without solving the equations~\eqref{eq:eq1}~and~\eqref{eq:eq2} by substituting the values of $\mu^{\alpha} = -E^{\alpha}_{ref}$ into Eq.~\eqref{eq:eq1_M}. The existence of absolutely stable solutions at temperature  $T$ is indicated by a negative value of   $I_T(-E^{\alpha}_{ref})$. Indeed, in the limit where $\mu^{\alpha}\to -\infty$ for all $\alpha$, $I_T$ is strictly positive. Consequently, if the function evaluates to a negative value at the point $\mu^{\alpha} = -E^{\alpha}_{ref}$, then there is a root lying somewhere between them.
\par
The Third Law of Thermodynamics dictates that a disordered phase cannot achieve absolute stability at  $T=0\K$~\cite{laughlin_third_2018, fedorov_third_2010, wagih_thermodynamics_2021}. Therefore, if a nanocrystalline phase is stable at some temperature, then there must exist a critical temperature $T_c$ below which absolute stability is lost. This transition confirms that the absolute stability of the nanocrystalline phase is fundamentally driven by entropic contributions to the free energy.

\subsection{Metastability Against Phase Separation}\label{subsec:nucleation}
Even lacking absolute stability, a segregated nanocrystalline state can remain metastable. When the thermodynamic driving force, governed by the Gibbs free energy of mixing, is insufficient for spinodal decomposition, phase separation proceeds via nucleation. Nucleation of a new phase requires overcoming an interfacial energy barrier, defined as the activation energy ($Q$). Driven by thermal fluctuations, this process is thermally activated and follows Arrhenius kinetics:
\begin{equation}
w\propto \exp\left[-\frac{Q}{kT}\right]\,.
\end{equation}

According to classical nucleation theory, the formation energy of a critical nucleus at GB~\cite{porter_phase_2021} is given by:

\begin{equation}\label{eq:Q}
Q = \frac{16\pi s}{3}\frac{\gamma_{int}^3}{G_{v}^2}\,,
\end{equation}

where $\gamma_{int}$ is the interfacial energy of the precipitate, $G_{v} = G_{nc}-G_{hull}$ is the chemical driving force -- the energy above the convex hull, and $s$ is a structural factor depending on the shape of the precipitate. By assuming mechanical stress relaxation at GBs and, thus, neglecting strain energy, Eq.~\eqref{eq:Q} reveals that the activation energy $Q$ depends quadratically on the free energy of mixing. Lowering $G_{nc}$ elevates $Q$, thereby slowing precipitation kinetics, leading to a prolonged incubation period or an elevated onset temperature for precipitate growth. Thus, by reducing the chemical potential and $G_{nc}$, the system raises the activation barrier for precipitation. This shifts the phase transformation threshold upward, providing kinetic stabilization even if absolute thermodynamic stability is not achieved.

\section{Cosegregation of Solutes with Identical Segregation Spectra}\label{sec:identical_coseg}

To isolate collective segregation phenomena from individual solute chemistry, we first evaluate an artificial system where all solute species share identical, synthesized segregation spectra. When comparing systems with different numbers of solute species ($m$), the total solute concentration is held strictly constant: $\sum_{\alpha} X^{\alpha}_{tot} = X_{tot}$. Consequently, any variations in segregation or thermal stability are driven exclusively by collective effects. Symmetry dictates that these collective interactions are maximized when individual solute concentrations are equal, yielding identical chemical potentials ($\mu^{\alpha} = \mu$ and $M^{\alpha} = M$ for all species $\alpha$). Under these symmetric conditions, the Mass Balance Equation~\eqref{eq:eq2}  is satisfied automatically, simplifying the Stability Equation~\eqref{eq:eq1} to:

\begin{equation}\label{eq:eq1_1d}
\sum_i\rho_i \ln(1+M\sum_\alpha \epsilon_i^{\alpha}) - \ln(1+mM) -\frac{\kgam}{kT} = 0\,.
\end{equation}
It is also useful to define total GB and inter-grain solute concentrations: $X_{gb} =\sum_{\alpha} X^{\alpha}_{gb}$ and $X_{c} =\sum_{\alpha} X^{\alpha}_{c}$.

\subsection{Competing Solutes}

In the case of competing solutes (the Pearson correlation coefficient for their segregation spectra $r=1$) $\epsilon_i^{\alpha} = \epsilon_i$ and Eq.~\eqref{eq:eq1_1d} reduces to:

\begin{equation}\label{eq:eq1_dsc}
\sum_i\rho_i \ln(1+m M \epsilon_i) - \ln(1+mM) -\frac{\kgam}{kT} = 0\,.
\end{equation}

It can be noted that by substituting $\overline{M} = m M$, Eq.~\eqref{eq:eq1_dsc} reduces to that for a single solute. Therefore, the root of Eq.~\eqref{eq:eq1_dsc} is $M{^{(m)}}=M{^{(1)}}/m$ (where the superscript $(m)$ denotes the number of solutes) and $\mu{^{(m)}}=\mu{^{(1)}}-kT\ln m$. Substituting these into Eq.~\eqref{eq:X} leads to $X_c^{\alpha}=X_c/m$ and $X^{\alpha}_i=X_i/m$. Ultimately, Eq.~\eqref{eq:fgb} gives ${f_{gb}}{^{(m)}} = {f_{gb}}{^{(1)}}$, indicating that the presence of a competing solute has no distinct influence on grain size stabilization. However, the change in chemical potential contributes to the Gibbs energy Eq.~\ref{eq:gibbs_mix}:

\begin{equation}
{G_{nc}}{^{(m)}}=  -kT\ln(1+mM{^{(m)}}) + \sum_{\alpha}{\mu^{\alpha}}{^{(m)}}X^{\alpha}_{tot} = -kT\ln(1+M{^{(1)}}) + \left(\mu{^{(1)}}-kT\ln m\right)X_{tot} = {G_{nc}}{^{(1)}} -kT X_{tot} \ln m\,.
\end{equation}

Thus, an effect identical to that observed in high-entropy alloys is present: multicomponent mixing introduces a logarithmic contribution to the free energy. That is, by mixing several solutes, the segregated state can be stabilized against precipitation, or the driving force for phase separation can at least be reduced. This phenomenon can be termed high-entropy stabilization.

\subsection{Cooperative Solutes}

\begin{figure}
    \centering
    \includegraphics[width=0.95\linewidth]{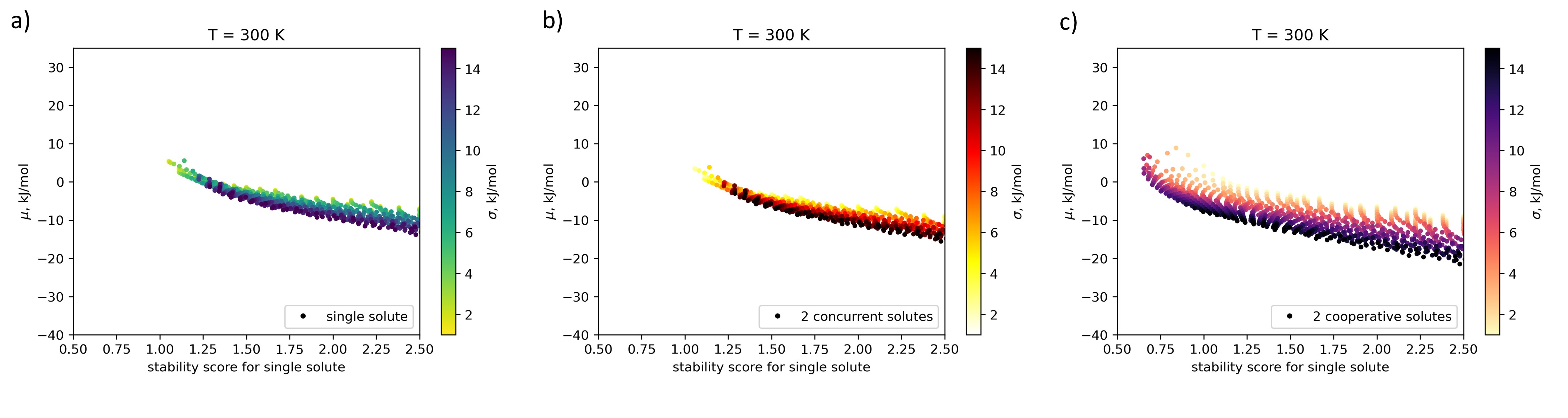}
    \caption{Chemical potential in the symmetric model at $300\K$ for Gaussian spectra with various parameters: (a) single solute, (b) two competing solutes, (c) two cooperating solutes.}
    \label{fig:mu_numeric}
\end{figure}

For cooperative solutes ($r=-1$) an elegant analytical solution is unavailable. To address this, numerical simulations of Eq.~\eqref{eq:eq1_1d} were performed across a range of spectra with varying mean energies, variances $\sigma$, and $\kgam$ values (Figure~\ref{fig:mu_numeric}). Given the correlation between the spectrum and $\kgam$ dependencies, it is convenient to plot $\mu$ against an integral characteristic: the stability score ($\Phi^M_{0K}$).
The residual data scatter is well captured by $\sigma$: specifically, an increased variance in segregation energies at a fixed stability score lowers the chemical potential in both single-solute and cosegregation regimes. This dependence is more pronounced under cooperative segregation. Although changing the temperature slightly alters the slope of the curves, this effect can be neglected in the first approximation. Additionally, elevated stability scores correspond to lower chemical potentials, confirming enhanced stability. Thus, cooperative solutes can form stable structures even when their individual stability scores fall below unity.
\par
\begin{figure}
    \centering
    \includegraphics[width=0.7\linewidth]{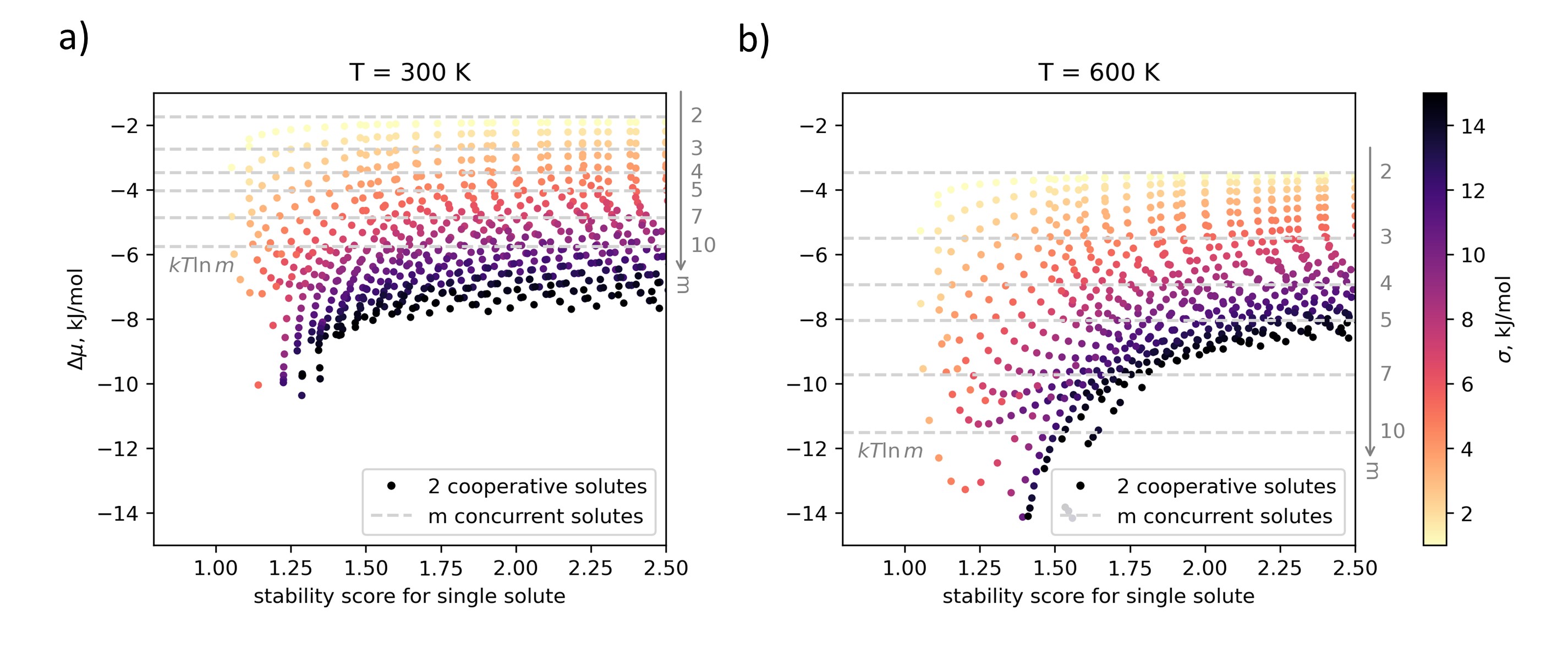}
    \caption{Change in chemical potential $\Delta \mu$ upon transitioning from a single solute to multiple solutes at $T=300\K$ (a) and $T=600\K$ (b). The dots denote the case of two cooperating solutes with the corresponding stability score and segregation energy variance $\sigma$, while gray dashed lines show $\Delta \mu$ for the case of $m$ competing solutes, which is independent of the spectrum shape and GB energy.}
    \label{fig:dmu}
\end{figure}

Figure~\ref{fig:dmu} displays the chemical potential shift $\Delta \mu$ accompanying the transition from single to multiple solutes.
Notably, the chemical potential reduction driven by two cooperating solutes always exceeds that for two competing solutes. While this reduction is more prominent in systems with lower stability scores (and higher initial $\mu$), the resulting  $\mu$ decreases monotonically with increasing  $\Phi_{0K}$ as can be seen from Figure~\ref{fig:mu_numeric}(b-c). Consequently, solutes featuring high $\Phi_{0K}$ values offer the greatest capacity for suppressing precipitation, indicating that this alloy screening can be restricted exclusively to candidates with individual   $\Phi_{0K} > 1$. Finally, it can be observed that elevating the temperature diminishes the relative advantage of cooperative chemical potential reduction over the competing mechanism.
\par
\begin{figure}
    \centering
    \includegraphics[width=0.5\linewidth]{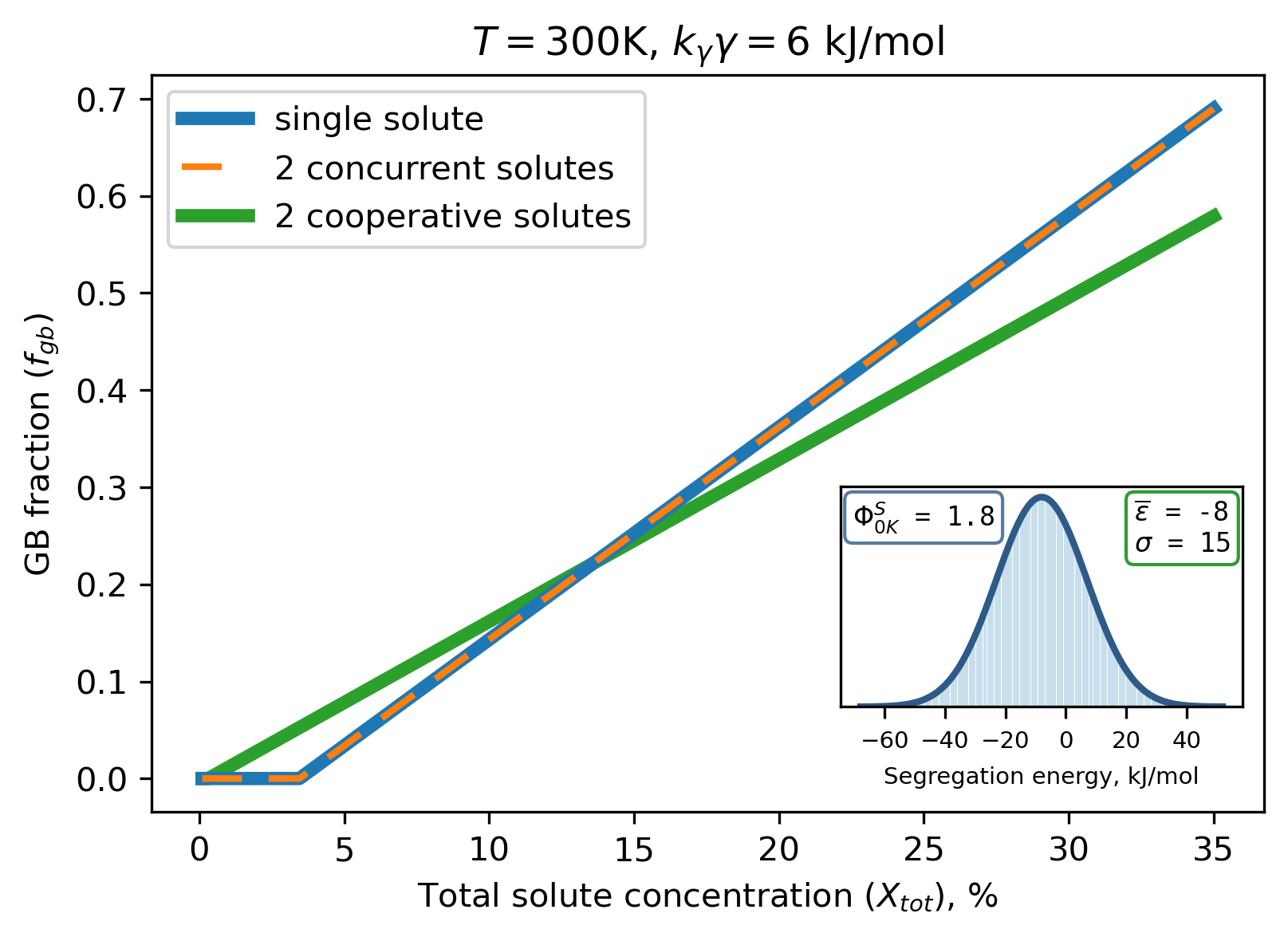}
    \caption{Grain boundary fraction as a function of total solute concentration compared across different cosegregation regimes at 300 K.  The inset displays the Gaussian segregation energy spectrum used for the model calculations. }
    \label{fig:fgb_sym}
\end{figure}

To evaluate grain size stabilization across different cosegregation modes, we first compare the solute concentrations within the grain at equal chemical potentials:

\begin{equation}
\left. {X^{\alpha}_c}(M)\right|_{\text{coop}} = \frac{M}{1+mM} = \left.X^{\alpha}_c(M)\right|_{\text{comp}}\,.
\end{equation}
Note that $\left.M\right|_{\text{coop}} < \left.M\right|_{\text{comp}}$ as mentioned earlier, which implies $\left.X^{\alpha}_c\right|_{\text{coop}} < \left.X^{\alpha}_c\right|_{\text{comp}}$. The value of $X_c$ represents the minimum total solute concentration required for stabilization, which must be sufficient to saturate the grains. Therefore, stabilizing a GB with a cooperative solute requires a smaller total amount of solute than in the case of competing solutes or single-solute segregation. This means that at higher temperatures, solutes desegregate less from the GBs, which aligns with the assumption of Wagih et al.~\cite{wagih_designing_2025}. However, numerical simulations demonstrate that under equilibrium, the GB often exhibits higher total solute concentration for cooperative solutes than for competing ones: $ \left. X_{gb}\right|_{\text{coop}} >\left. X_{gb}\right|_{\text{comp}}$. This implies that a higher total solute content is required to stabilize the GB at the given grain size when $X_{tot} \gg X_c$. Therefore, cooperative solutes suppress grain growth more effectively in the low-concentration, coarse-grain regime ($X_{tot}\ll X_c$), whereas they do so less effectively in the high-concentration, fine-grain regime (Figure~\ref{fig:fgb_sym}). This behavior offers a distinct advantage in alloy design, provided that grain growth is primarily driven by temperature-driven desegregation~\cite{jiao_nanocrystalline_2018}. 
\subsection{Intermediate Correlation Regimes}
\begin{figure}
    \centering
    \includegraphics[width=0.5\linewidth]{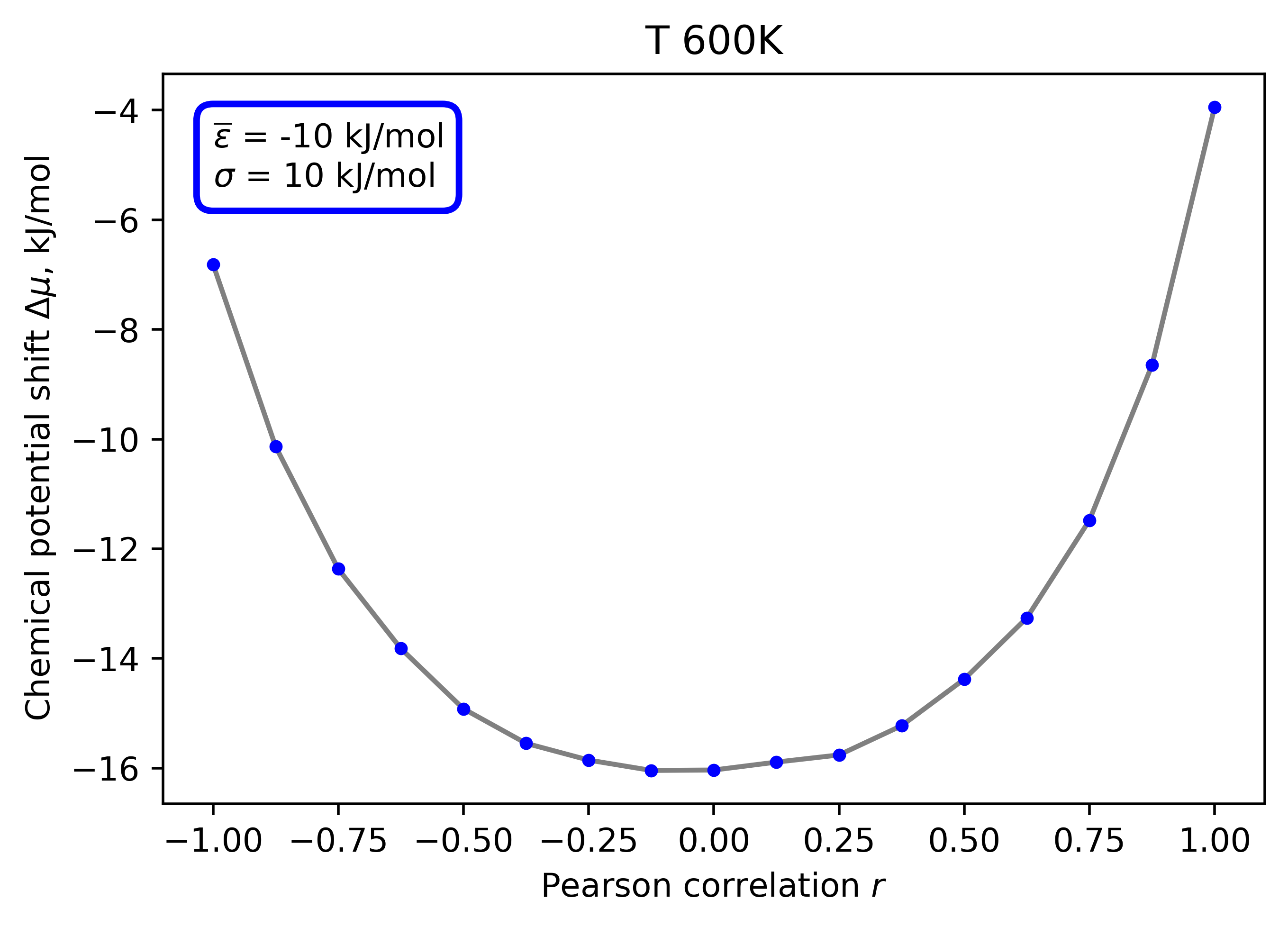}
    \caption{Chemical potential shift as a function of the correlation coefficient $r$ for cosegregation of two solutes with Gaussian segregation spectrum with mean segregation energy $\overline{\varepsilon}=-10$ kJ/mol and variance $\sigma=10$ kJ/mol at 600 K.}
    \label{fig:dmu_vs_r}
\end{figure}

For intermediate correlations ($-1<r<1$) numerical solutions of Eq.~\eqref{eq:eq1_1d} were performed, showing that the minimum of the chemical potential is achieved when the segregation spectra are completely uncorrelated ($r=0$). An example of the chemical potential dependence on $r$ is shown in Figure~\ref{fig:dmu_vs_r}. It can be seen that as $r$ approaches $0$, the chemical potential decreases significantly, and at the optimum the reduction in chemical potential is even greater than the sum of the effects at $r=1$ and $r=-1$. At present, the exact reason for this system behavior is not obvious. Apparently, the combined action of entropic ($r=1$) and energetic ($r=-1$) effects leads to synergy. Consequently, the cosegregation of solutes with uncorrelated spectra stabilizes the nanocrystalline phase most effectively. This result is quite unexpected, since it was previously implicitly assumed that the most interesting effects are achieved when the spectra are anti-correlated ($r=-1$).
\section{Atomistic Simulation Methods}\label{sec:methods}
We apply the developed theory to search for substitutional solutes that stabilize nanocrystalline silver. To model the cosegregation of real solutes, it is necessary to calculate the model parameters: segregation energy spectra, structural GB energy and reference energies of solutes. 
\par
Calculations were automated using a Python workflow.  Atomistic calculations were performed using GPUMD~\cite{gpumd} with the foundation model interatomic potential NEP89~\cite{nep89}. This potential provides a reasonable computational cost and quantum-level accuracy for GB and segregation calculations involving Ag and 15 other species (Al, Au, Cr, Cu, Mg, Mo, Ni, Pb, Pd, Pt, Ta, Ti, V, W, Zr). Furthermore, it enables the calculation of segregation for other species, albeit with lower accuracy. OVITO~\cite{stukowski_visualization_2010} was used for visualization and identification of GB atoms. Machine learning (ML) methods, in particular principal component analysis (PCA), k-means clustering and linear regression were performed using the Python scikit-learn package~\cite{scikit}.
\par
A $15\times15\times15~\text{nm}$ silver polycrystal containing $10$ grains was generated via the Voronoi tessellation method in the Atomsk package~\cite{hirel_atomsk_2015}. Following initial energy minimization using the FIRE algorithm~\cite{fire}, which ensures structural stability, the sample underwent MD annealing at 500 K for 500 ps under a Berendsen thermostat and barostat~\cite{berendsen_molecular_1984} to equilibrate the GB structure. A final energy minimization was performed before the calculation of segregation spectra to eliminate thermal atomic vibrations.
\par 
Using the common neighbor analysis algorithm, atoms belonging to GB were identified in the generated structure. A constant cutoff radius of $3.5$~\AA  \ was used. The segregation energy of a solute $\alpha$ at site $i$ ($E^{\alpha}_{i}$) was calculated as the energy difference between the total energy of the relaxed system with the solute at this site, $E_{sys}({\alpha}, i)$, and that with the solute placed in the bulk away from all defects, $E_{sys}({\alpha},c)$: $E^{\alpha}_{i} =  E_{sys}({\alpha},i) - E_{sys}({\alpha},c)$.
\par
Computing a complete segregation spectrum on an Nvidia RTX 3080~Ti GPU requires approximately 83~hours. To mitigate this substantial computational cost, we adopted the ML approach proposed by Wagih et al.~\cite{wagih_learning_2020}, which drastically reduced the computational time to just 2~hours by predicting the complete spectrum based on the calculation of only 200~selected points.  To check the model for overfitting, the dataset was divided equally into training and testing subsets.  The training of the ML models relies on invariant descriptors of the local atomic environment rather than raw coordinates, ensuring invariance under translations, rotations and permutations of identical atomic species. For GB atoms, SOAP descriptors~\cite{bartok_representing_2013} were calculated using the DScribe library~\cite{dscribe} with a cutoff radius $R_{\text{cut}}$ of $6$~\AA, to encompass several coordination spheres and the number of radial and angular basis functions $n_{\text{max}} = 12$ and $l_{\text{max}} = 12$, respectively. To prevent overfitting while maintaining the advantages of a concise training set, PCA method was applied; the first 20 principal components were found to explain 99\% of the data variance. Representative atomic sites covering the full diversity of local structural configurations were then extracted from the descriptor space via k-means clustering. Finally, a linear regression model was trained on the segregation energies computed for the selected sites and their corresponding descriptors. The model accuracy, validated on the remaining half of the dataset, yielded a mean absolute error (MAE) of 4.8 kJ/mol. While this accuracy is sufficient for high-throughput screening and qualitative trend analysis, comparison with experimental data would require expanding the training dataset and potentially deploying non-linear regression architectures.
\par
Structural GB energy was calculated as the per atom energy difference between the polycrystal and a $10\times10\times10 $ unit cells single crystal, divided by the GB fraction of the polycrystal:

\begin{equation}
\kgam = \frac{E_{poly}/N_{poly}-E_{single}/N_{single}}{f_{gb}}\,,
\end{equation}

where $E_{poly}$ and $N_{poly}$ denote the energy and number of atoms in the polycrystal, respectively, while $E_{single}$ and $N_{single}$ represent the energy and number of atoms in the single crystal.
\par
To calculate the reference energy, it is necessary to determine both $E_{sol}^{\alpha/A}$ and the formation energy of the intermetallic compound with the lowest solute stoichiometry $x_l$ ($E_{form}(A_{1-x_l}\alpha_{x_l})$). The energy of embedding one solute atom $\alpha$ into the matrix $A$ was calculated as $E_{sol}^{\alpha/A}= E(A_{N-1}\alpha) - ((N-1)E_{coh}^A + E_{coh}^{\alpha})$, where $E( A_{N-1}\alpha)$ is the energy of a $10\times10\times10 $ unit-cell single crystal, with one atom replaced by an atom $\alpha$, and $E_{coh}^{\alpha}$ is the cohesive energy, calculated for a single crystal element $\alpha$ with dimensions of $10\times10\times10 $ unit cells. The intermetallic compound with the lowest stoichiometry and its crystallographic structure were determined based on the Materials Project database~\cite{mp1, mp2, mp_phase}. To ensure self-consistency, we recalculated the formation energies of these intermetallic compounds using the same NEP89 potential~\cite{nep89}: $E_{form}(A_{1-x_l}\alpha_{x_l})= E(A_{1-x_l}\alpha_{x_l}) - (1-x_l)E_{coh}^A - x_lE_{coh}^{\alpha}$. However, to improve reliability, the condition in Eq.~\eqref{eq:multi_inter_cond} was verified using formation energies taken from the Materials Project database.
\par
The current model of cosegregation does not account for solute--solute interactions, and therefore the results may differ significantly for some systems in which solute--solute interactions are strong. For example, strong attractive solute--solute interactions significantly increase the stability score of Ag(Ni)~\cite{marchiy_spectral_2025}, so the resulting stability scores for such systems will be underestimated. Conversely, systems characterized by a negative mixing energy typically exhibit strong repulsive interactions, as these solute atoms prefer matrix neighbors over like atoms, thereby leading to overestimated stability scores. Furthermore, interactions between distinct solute species can either promote or suppress segregation~\cite{wagih_thermodynamics_2021}.

\section{Cosegregation of Different Solutes in Silver}\label{sec:results}

All parameters required for the modeling, including segregation spectra, reference energies, and structural GB energy $\kgam = 9.32\text{ kJ/mol}$, were obtained using atomistic calculations.  In this section, the synergistic effects of multicomponent GB segregation are analyzed using specific examples. First, we examine how the range of stabilizing solutes expands by  including pairs that suppress grain growth together, even if they fail to do so individually. Second, we demonstrate that solutes capable of suppressing grain growth individually, but prone to phase separation, become resistant to phase separation upon cosegregation.

\subsection{The Emergence of Synergetic Stabilization of Grain Growth}\label{subsec:syn_stability}
\begin{figure}
    \centering
    \includegraphics[width=0.75\linewidth]{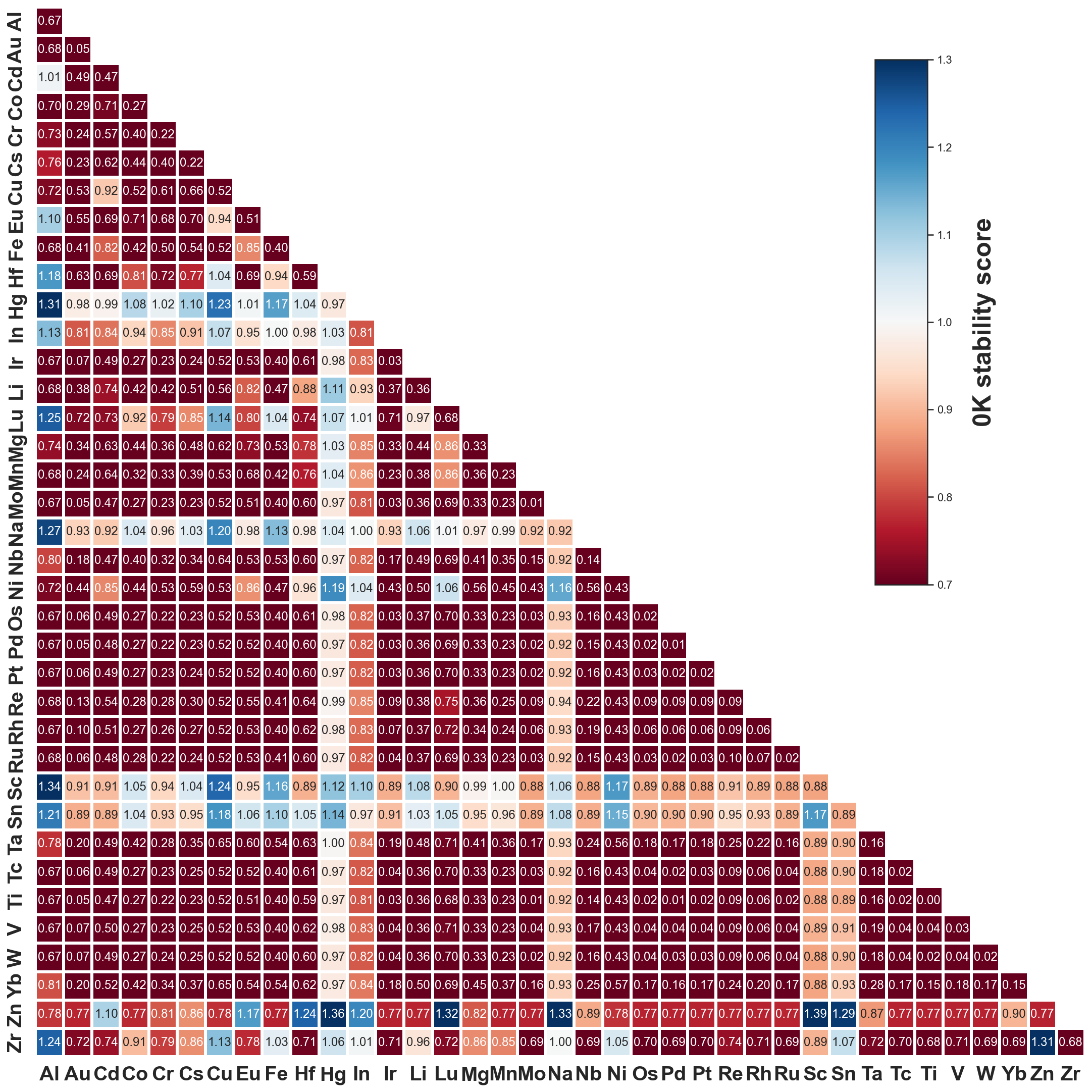}
    \caption{0 K stability score for solute pairs (the diagonal shows the scores for a single solute, while the off-diagonal cells represent pairs).}
    \label{fig:synergetic_stability_scan}
\end{figure}

As detailed in Section~\ref{subsec:phi_M_0K}, cosegregation invariably enhances the stability score $\Phi_{0K}^{M}$. Figure~\ref{fig:synergetic_stability_scan} highlights solute pairs that do not individually suppress grain growth ($\Phi^M_{0K}(\alpha)<1$ and $\Phi^M_{0K}(\beta)<1$), yet they exhibit synergistic stabilization: $\Phi^M_{0K}(\alpha, \beta)>1$. It is worth noting that the stability scores tend to be underestimated for systems featuring attractive interactions, which are especially strong in Ag(Ni), Ag(Cu) to name a few.
\par
\begin{figure}
    \centering
    \includegraphics[width=1\linewidth]{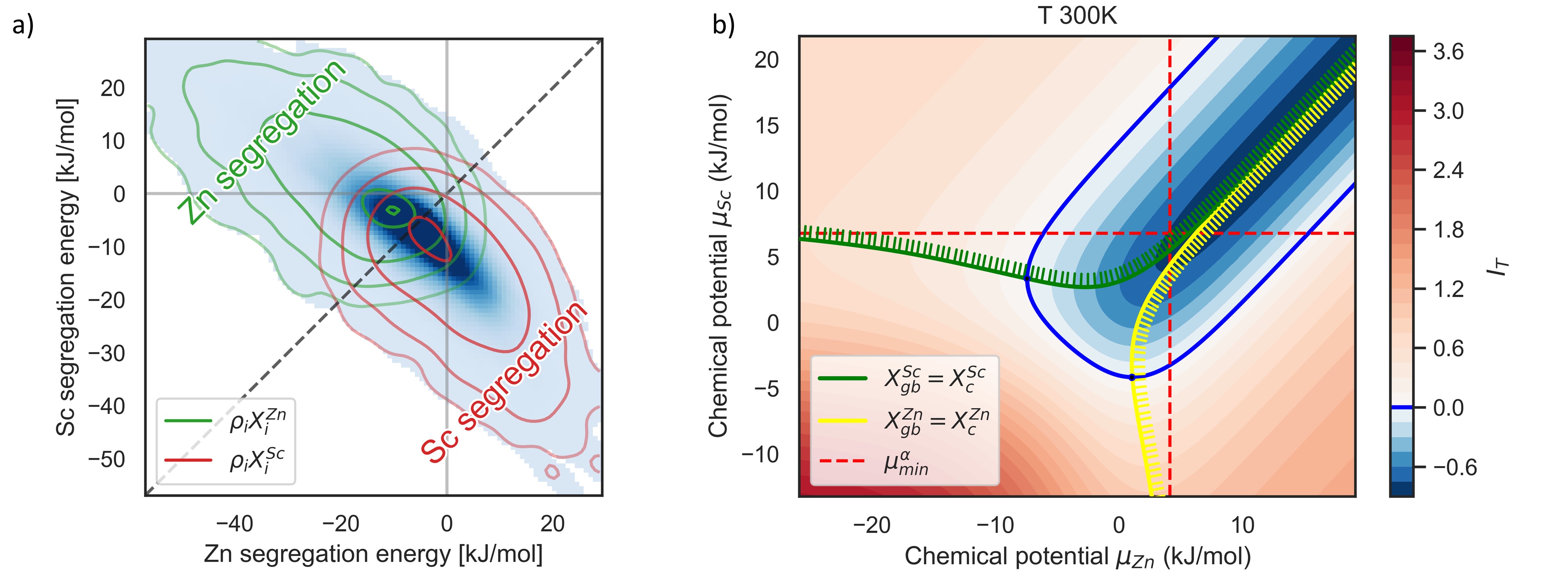}
    \caption{Thermodynamics of cosegregation in the Ag(Zn, Sc) system at $300\K$. (a) Joint probability density (bivariate segregation spectrum). The color gradient scales with the probability density, where the highest intensities represent maximum probability. Isolines indicate the individual distributions of Zn (green) and Sc (red) evaluated at the chemical potentials corresponding to the lowest total concentration required for GB stabilization.  (b) $I_T(\{\mu^{\alpha}\})$ function values. Red dashed lines mark the individual minima of $I_T(\{\mu^{\alpha}\})$ asymptotes for each solute separately. Green and yellow lines highlight the $X_{gb}^{\alpha}=X_c^{\alpha}$; on their shaded side the solute desegregation occurs ($X_{gb}^{\alpha}<X_c^{\alpha}$), destroying the nanocrystalline stability. The solid blue line represents the roots of Eq.~\eqref{eq:eq1}, of which only those on the unshaded side between the green and yellow boundaries are physically meaningful. }
    \label{fig:Coseg_Zn_Sc}
\end{figure}

The most stable collective performance occurs in the Zn--Sc pair, which exhibits $\Phi_{0\text{K}}^{M} = 1.39$; therefore, we simulate this system in detail. Note that the segregation spectra in this pair exhibit pronounced anti-correlation: $r=-0.8$ (see Figure~\ref{fig:Coseg_Zn_Sc}(a)), which allows for a substantial reduction in their collective stability score.
\par
Solving Eq.~\eqref{eq:eq1} can be difficult in multidimensional cases, as the success of root-finding algorithms strictly depends on the quality of the initial guesses. Therefore, to find the roots of Eq.~\eqref{eq:eq1}, the following considerations are helpful. For each solute individually, there exists a chemical potential $\mu_{min}^{\alpha}$ where $X_{gb}^{\alpha}=X_{c}^{\alpha}$ is reached and the minimum of function $I_T(\mu^{\alpha})$ is achieved (red dashed lines in Figure~\ref{fig:Coseg_Zn_Sc}(b)). All $\mu^{\alpha}>\mu_{min}^{\alpha}$ correspond to $X_{gb}^{\alpha}<X_{c}^{\alpha}$ and describe a case of desegregation. Segregation and the nanocrystal stability for a single solute can only be present at $\mu^{\alpha}<\mu_{min}^{\alpha}$. In the case of cosegregation, the chemical potentials where $X_{gb}^{\alpha}<X_{c}^{\alpha}$ shift, but tend towards their individual values in the limit $\mu^{\beta}\to-\infty $  (shown by green and yellow lines in Figure~\ref{fig:Coseg_Zn_Sc}(b)). These asymptotes provide excellent initial approximations for identifying such chemical potentials in the two-dimensional case. By tracing along the  $X_{gb}^{\alpha}=X_{c}^{\alpha}$ curves, we can find two roots of $I_T$. These points delimit the solution domain and provide initial guesses for solving Eq.~\eqref{eq:eq1}; regions where solutions cannot exist are marked with green and yellow hatching in Figure~\ref{fig:Coseg_Zn_Sc}(b).
\par
\begin{figure}
    \centering
    \includegraphics[width=0.5\linewidth]{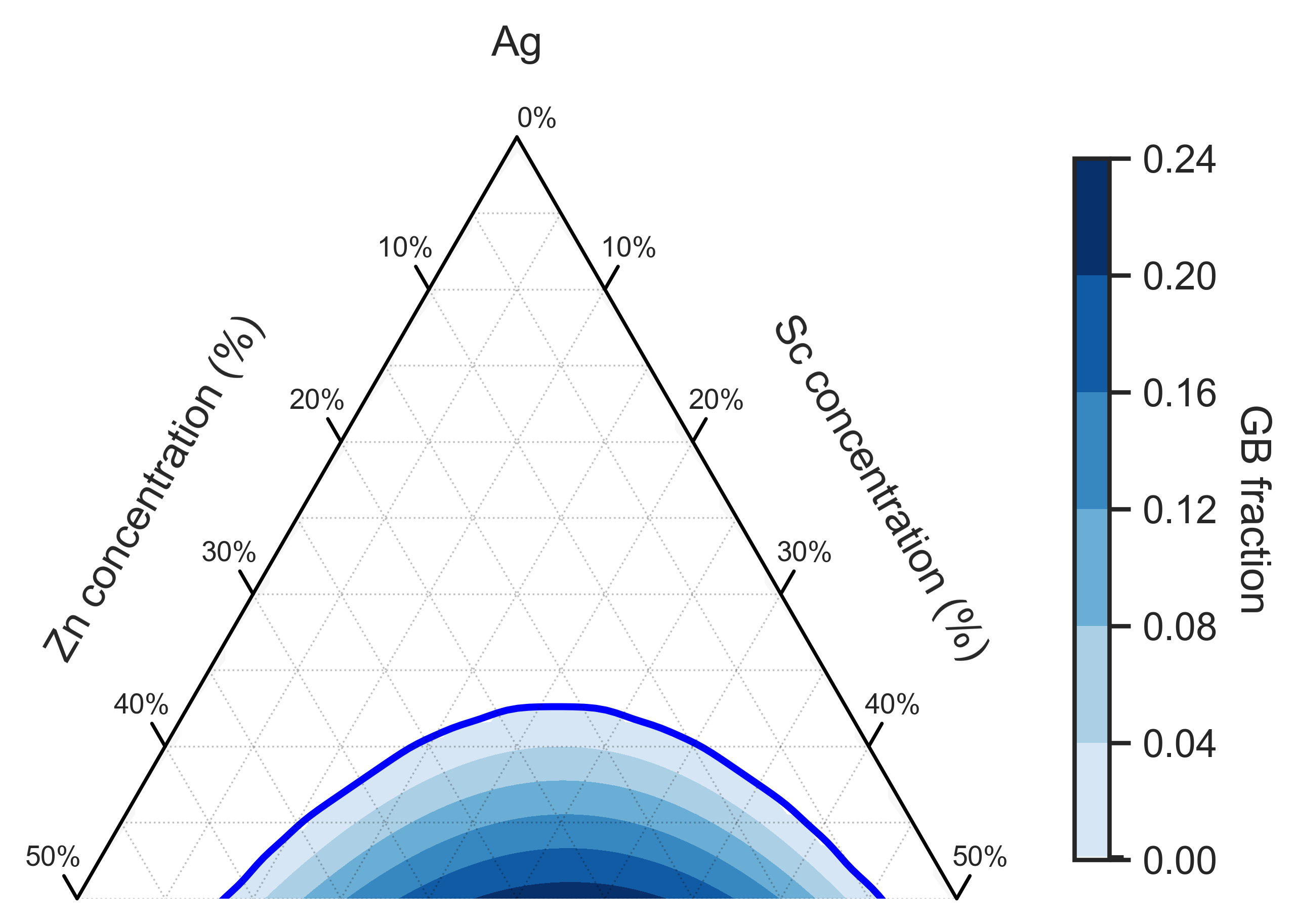}
    \caption{Ternary diagram for equilibrium grain boundary fraction in Ag(Zn, Sc) at $T=300\K$.}
    \label{fig:fgb_Zn_Sc}
\end{figure}

Using the dependence of $\mu^{Sc}$ on $\mu^{Zn}$ to solve Eq.~\eqref{eq:eq2}  provides the relation between $f_{gb}$ and the total solute concentrations (see Figure~\ref{fig:fgb_Zn_Sc}). Although valid solutions with $f_{gb}>0$ exist, they demand an exceptionally high total solute content exceeding $35\%$ even at room temperature. Raising the temperature further elevates the minimum required solute concentration. This behavior stems from the low segregation energies of both solutes, which promote desegregation away from GBs. Such weak segregation necessitates a high solute concentration to achieve the GB saturation required for stabilization. It should be noted that the model neglects solute--solute interactions, and therefore its applicability in such a high--concentration case is limited. 
\par
Thus, cosegregation of solutes that fail to stabilize the nanocrystalline phase individually provides minimal benefit from the standpoint of stabilization. Even the solute pair with the highest collective stability score requires an exceptionally high total solute concentration to suppress grain growth.

\subsection{Stabilization Against Phase Separation}\label{subsec:abs_stability}

\begin{figure}
    \centering
    \includegraphics[width=1\linewidth]{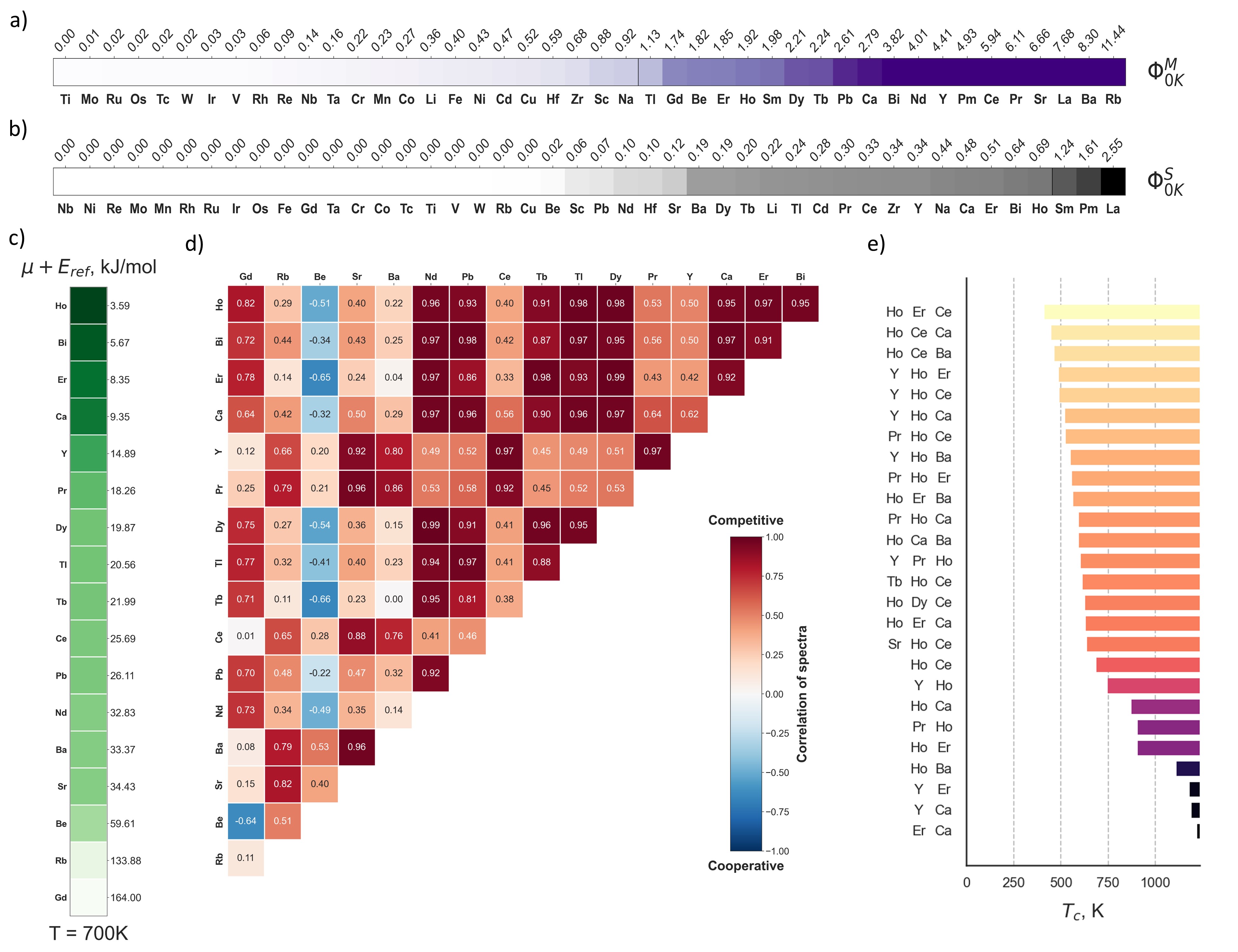}
    \caption{Solute segregation stabilization screening:  (a) Metastability and (b) absolute stability scores at 0 K. (c) Chemical potential relative to phase-separated states $\mu+E_{ref}$ at $700\K$. (d) Pearson correlation coefficients for the segregation spectra of solutes stabilizing grain growth at $700\K$. (e) Temperature range of phase-separation stability for the considered solute pairs and triplets. For triplets, only those providing a stabilization temperature below that of the best solute pair are shown. More stable compositions, which exhibit a lower $T_{c}$, are indicated by lighter colors.}
    \label{fig:sreening}
\end{figure}

More promising effects occur when cosegregating solutes that individually suppress grain growth but are susceptible to phase separation. Figures~\ref{fig:sreening}(a-b) illustrate calculated respective individual metastability scores $\Phi^M_{0K}$ and absolute stability scores~\cite{wagih_thermodynamics_2021}:

\begin{equation}
\Phi^S_{0K}(\alpha)=-\frac{1}{k_\gamma \gamma} \sum_{i}^{E^{\alpha}_i<E^{\alpha}_{ref}} \rho_i(E^{\alpha}_i-E^{\alpha}_{ref})\,.
\end{equation}

If the corresponding score exceeds one, the nanocrystalline phase with segregated solute is metastable or absolutely stable at 0 K. Figure~\ref{fig:sreening}(b) reveals that segregation of La, Sm and Pm yields  $\Phi_{0\K}^{S}>1$. This violates the Third Law of Thermodynamics, since only ordered phases~\cite{laughlin_third_2018, fedorov_third_2010} are stable at 0 K. The omission of repulsive solute--solute interactions, which increase the energy of the segregated state~\cite{wagih_thermodynamics_2021}, likely explains this discrepancy. Therefore, these elements were excluded from the analysis. Conversely, the stability scores of some individual solutes may be underestimated, as already noted in Section~\ref{subsec:syn_stability}. Additionally, certain solutes~(Al,  Au, Cs, Eu, Hg, In, Lu, Mg, Pd, Sn, Tm, Zn) possessed positive reference energies, indicating a failure of the potential to accurately reproduce different phase energies for those species; they were therefore also excluded from the analysis.
\par
To determine the proximity of  individual solutes to absolute stability, chemical potentials relative to the precipitating phase $\mu+E_{ref}$ at $T=700\K$ were computed for solutes possessing metastability at this temperature (see Figure~\ref{fig:sreening}(c)). Because these chemical potentials exhibit weak temperature dependence, this solute ranking remains virtually unchanged. No solutes achieve absolute stability at this temperature, since $\mu+E_{ref}>0$ for all cases.  Notably, while $\Phi_{0K}^{S}$ values and $\mu+E_{ref}$ magnitudes generally correlate, this relationship breaks down for certain elements. For example, the segregation of Na, Zr, Cd, Li fails to establish metastability, whereas elements with lower $\Phi_{0K}^{S}$ values can suppress grain growth.  Since these anomalies arise even among effective grain-boundary stabilizers, assessment of absolute stability via the $\mu+E_{ref}$ is more reliable.

\par
In Figure~\ref{fig:sreening}(d), solutes were sorted in descending order of $\mu+E_{ref}$ values, meaning that solute pairs with the greatest potential for stability are located in the right part of the table. While the four most stable solutes (Ho, Bi, Er, Ca) exhibit strongly correlated spectra, the next two solutes (Y and Pr) show a weaker correlation with the former group, yielding a Pearson correlation coefficient of $r\approx 0.5$. 
\par
As detailed in Section~\ref{subsec:abs_stability_criterion}, the absolute stability of any solute combination satisfying Eq.~\eqref{eq:multi_inter_cond} can be verified by checking the sign of $I_{T}(-E^{\alpha}_{ref})$, where a negative sign confirms absolute stability. Figure~\ref{fig:sreening}(e) illustrates the critical absolute stabilization temperatures $T_c$ for the identified solute pairs and triplets. To avoid cluttering the plot, only the triplets yielding a lower $T_c$ than the best solute pair are presented.
\par
It is useful to consider the GB solute concentration corresponding to absolute stability. We found that it generally decreases slightly with temperature. This decrease is driven by the drop of the optimal value of $\mu^{\alpha}$ with temperature elevation; however, the concurrent desegregation generally diminishes this effect rendering the temperature dependence of the GB concentration rather weak. Therefore, the GB concentration at $\mu^{\alpha} = -E_{ref}^{\alpha}$ and $T=T_c$ provides a useful estimate. Across all identified pairs and triplets, the total GB solute concentration reaches approximately $50\%$ to $60\%$. This high equilibrium occupancy underscores the necessity of accounting for solute-solute interactions in future high-fidelity models. Therefore, we focus here on the qualitative results obtained without them to clearly emphasize the role of spectral characteristics alone.
\par
Analysis of the solute pairs reveals that Ho--Ce and Ho--Y possess the greatest stability as a result of Ho's standalone proximity to stability and their low spectral correlation coefficients ($r= 0.4$ and $r= 0.5$, respectively). Although Pr features a higher individual proximity to stability than Ce and nearly matches Y, the Ho--Pr pair ($r= 0.53$) proved less stable, illustrating the powerful role of spectral correlation. All mentioned cases involve a mixture of entropic and energetic effects of cosegregation. Under purely entropic conditions, the Ho--Bi pair is expected to perform best, but the formation of the stable intermetallic compound BiHo makes this difficult to assess, since the criterion in Eq.~\eqref{eq:BC_inter_cond} is violated. Among the entropy-driven pairs satisfying Eq.~\eqref{eq:BC_inter_cond} Ho--Ca ($r= 0.95$) dominates, followed by Ho--Er ($r= 0.97$). Because Ca succeeds over Er despite a $1\ \text{kJ/mol}$ lower individual proximity to stability, we can conclude that a minor $2\%$ decrease in spectral correlation is more impactful than a $1\ \text{kJ/mol}$ chemical potential difference.
\par
Incorporating a larger number of cosegregating solutes increases configuration variability and strengthens the entropic effect. Consequently, certain solute triplets yield lower stabilization temperatures than all solute pairs.  Among triplets, the leading combination is Ho--Er--Ce, which exhibits a critical stabilization temperature of $415\K$. This triplet consists of competing (Ho, Er) and partially complementary (Ce) solutes. The following are combinations in which Er and Ce are replaced by Ca, Ba, Y, Pr, Tb, Dy or Sr. All compounds without Ho have a $T_c$ higher than that of the leading solute pair (Ho--Ce), and are therefore not marked in Figure~\ref{fig:sreening}.
\par
Solute combinations involving Be, Rb, and Gd are absent from the leading pairs and triplets, since their standalone distances from stability  --- $60$, $134$ and $164\ \text{kJ/mol}$,  respectively -- are too large for cosegregation to counteract them.
\par
\begin{figure}
    \centering
    \includegraphics[width=1\linewidth]{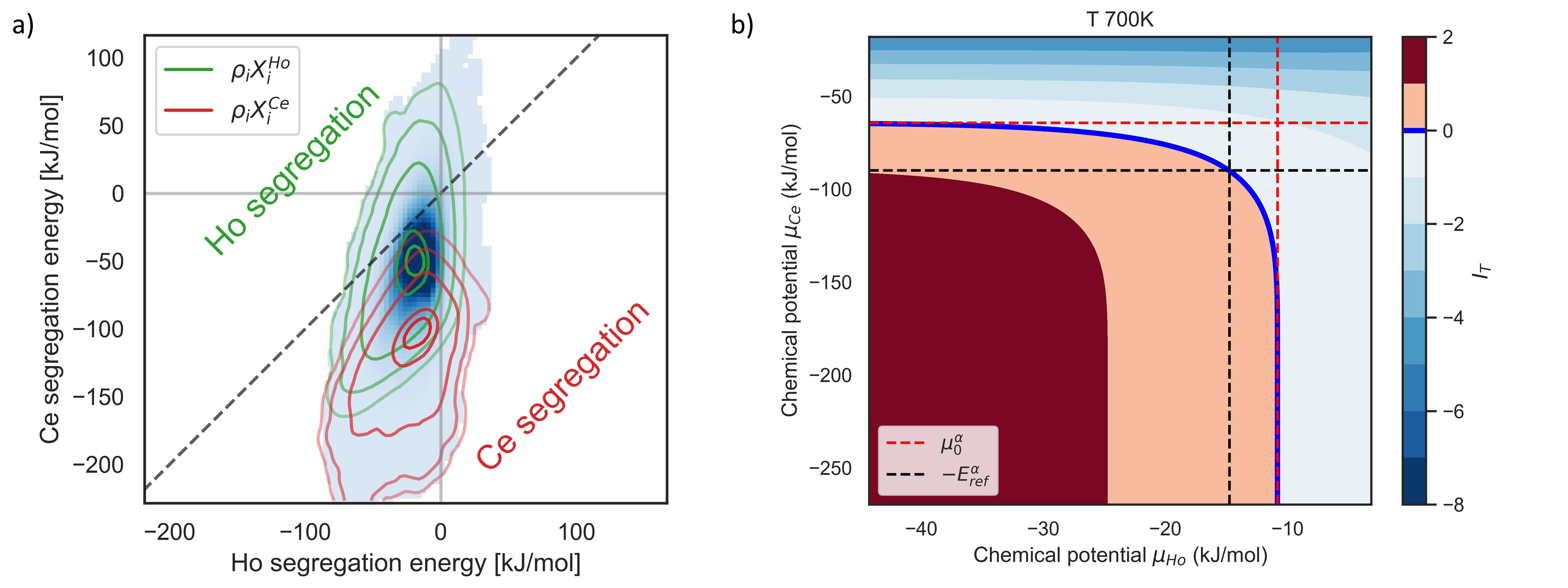}
    \caption{Thermodynamics of cosegregation in the Ag(Ho, Ce) system at $700\K$. (a) Joint probability density (bivariate segregation spectrum). The color gradient scales with the probability density, where the highest intensities represent maximum probability. Isolines indicate the individual distributions of Ho (green) and Ce (red) evaluated at the chemical potentials corresponding to the lowest total concentration required for GB stabilization.  (b) $I_T(\{\mu^{\alpha}\})$ function values. Red dashed lines mark the asymptotes of the individual solutions, while black dashed lines indicate the level below which the chemical potential corresponds to absolute stability. The blue solid line marks the roots of Eq.~\eqref{eq:eq1}.}
    \label{fig:Coseg_Ho_Ce}
\end{figure}

Moving forward, we transition from this broad screening to the detailed modeling of specific systems. For simplicity, we restrict this analysis to binary cosegregation; although extending the model to multicomponent systems is straightforward, it is computationally expensive. We focus on the Ho--Ce system, where the solute spectra are weakly correlated $r=0.4$ (see Figure~\ref{fig:Coseg_Ho_Ce}(a)). This low correlation triggers both cooperative and concurrent cosegregation behavior. Furthermore, since both solutes also possess high individual stability, their cosegregation successfully induces absolute stability starting from just 700 K. Figure~\ref{fig:Coseg_Ho_Ce}(a) shows that there are both regions of overlap in solute distributions, which increase entropy, and regions of individual segregation for each solute, which decrease the segregation energy.
\par 
To establish initial approximations for solving Eq.~\eqref{eq:eq1}, we note that each solute element independently stabilizes the metastable nanocrystalline phase. Consequently, when $\mu^{Ce}\to-\infty$, Eq. \eqref{eq:eq1} reduces to the single-element one with root $\mu^{Ho}_0$ (and vice versa). These asymptotes --- $\mu^{Ho}_0$, $\mu^{Ce}_0$ --- serve as effective starting values for the numerical solver. Notably, the collective solution diverges most significantly from the individual ones near the point of asymptote intersection. In this region  $\mu^{\alpha}$ decreases as a result of the entropic (competitive) or energetic (cooperative) cosegregation effects revealed in Section~\ref{sec:identical_coseg} (see Figure~\ref{fig:Coseg_Ho_Ce}(b)). These outcomes clearly demonstrate that at $700\K$, roots emerge where $\mu^{\alpha}$ drops enough to shift $\mu^{\alpha}+E^{\alpha}_{ref}$  below zero.
\par
\begin{figure}
    \centering
    \includegraphics[width=1\linewidth]{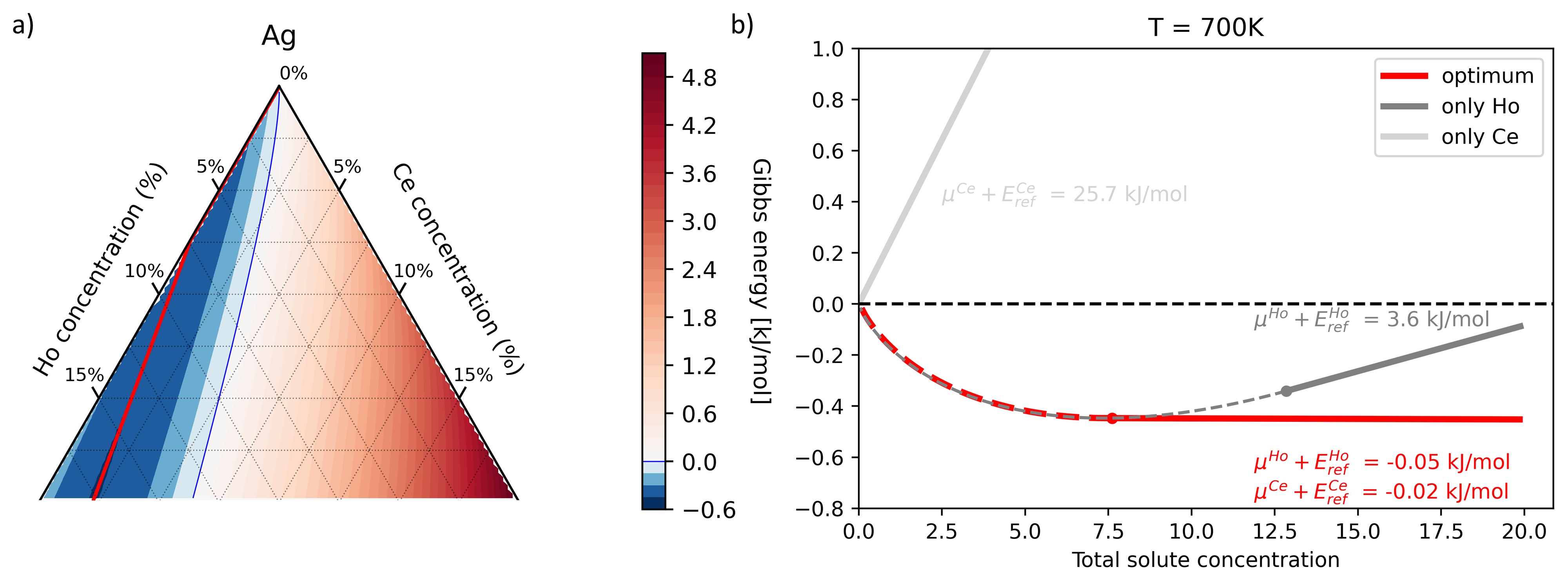}
    \caption{Thermodynamics of the Ag(Ho, Ce) system at $700 \K$:  (a) Ternary phase diagram for Gibbs free energy of the silver-rich phase, where the red line represents the lowest energy states in the sections of constant total solute concentration. (b) Comparison of the Gibbs free energy at the optimal concentration ratio against the energies of single-solute segregation systems as a function of total solute concentration. Dashed lines indicate the bulk solid solution where nanocrystalline stability is absent, whereas dots denote the onset of nanocrystalline stabilization, acting as boundary markers that separate the dotted and solid curves. }
    \label{fig:Gibbs_Ho_Ce}
\end{figure}

Finally, solving the system of equations~\eqref{eq:eq1},~\eqref{eq:eq2} we obtain the phase diagram shown in Figure~\ref{fig:Gibbs_Ho_Ce}(a). The Ho-to-Ce concentration ratio that minimizes the Gibbs energy (indicated by the red line in  Figure~\ref{fig:Gibbs_Ho_Ce}(a)) follows a linear relationship, $X^{Ho}_{tot} = aX^{Ce}_{tot} + b$, where coefficients $a$ and $b$ are temperature-dependent. Along this optimum trajectory, the chemical potentials of the solutes (whose sum is equal to the slope of the free energy curve on the plot)  remain constant. Notably, the Gibbs energy at this optimal concentration ratio is lower than that of either constituent solute acting individually (see Figure~\ref{fig:Gibbs_Ho_Ce}(b)). The data reveal a trade-off: Ho segregation exhibits higher stability against precipitation but is less efficient at suppressing grain growth (requiring $X_{tot}^{Ho}$ at least $\sim13\%$ for stabilization), whereas Ce segregation is more prone to precipitation but highly effective at suppressing grain growth (requiring $X_{tot}^{Ce}$ only at least $\sim 0.002\%$). Combining both solutes averages the thermal stabilization effect, but creates a mixture that resists precipitation better than either element alone. However, it is worth noting that even at a low total solute concentration, the GB concentration can be remarkably high. Indeed, in the optimal state, the GB is filled with $49\%$ Ho and $9\%$ Ce, reaching a total occupancy of $58\%$, which suggests the relevance of accounting for solute-solute interactions.
\par
\begin{figure}
    \centering
    \includegraphics[width=0.5\linewidth]{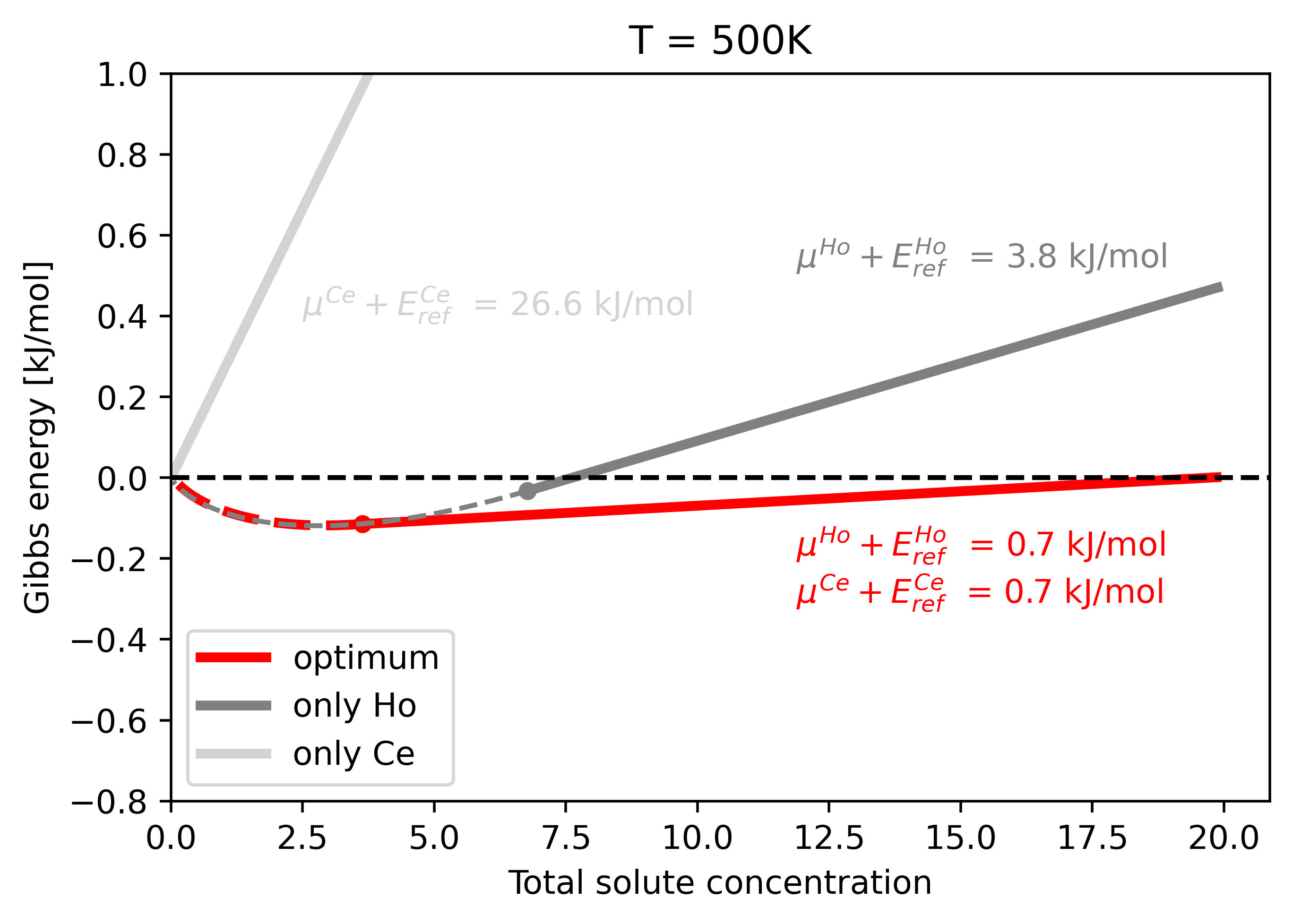}
    \caption{Comparison of the Gibbs free energy in the Ag(Ho, Ce) system at $500\K$ at the optimal concentration ratio against the energies of single-solute segregation systems as a function of total solute concentration. Dashed lines indicate the bulk solid solution where nanocrystalline stability is absent, whereas dots denote the onset of nanocrystalline stabilization, acting as boundary markers that separate the dotted and solid curves.}
    \label{fig:G1d_Ho_Ce_500K}
\end{figure}

 
Consider the same system at $T=500\K$. Because the temperature is below $T_c$ the system is prone to phase separation. However, the driving force for precipitation under cosegregation is significantly smaller: the $\mu^{\alpha}+E_{ref}^{\alpha}$ value drops by a factor of $5.4$ compared to Ho segregation alone, and by a factor of $38$ compared to Ce segregation (see Figure~\ref{fig:G1d_Ho_Ce_500K}). The activation energy $Q$ for precipitation (Eq.~\eqref{eq:Q}) during cosegregation increases by a factor of $30$ compared to the Ag(Ho) system and by a factor of $1444$ compared to the Ag(Ce) system.
Consequently, solute cosegregation can effectively suppress phase separation even at temperatures below $T_c$.
\section{Conclusion}\label{sec:conclusion}
We have developed a thermodynamic framework for multi-solute grain boundary (GB) cosegregation in polycrystals that captures the spectral nature of GBs while neglecting solute--solute interactions. The derived equations describe the emergence of nanocrystalline stability upon segregation and establish a framework to quantify stability with respect to phase separation.
\par
Our model demonstrates that the primary benefit of cosegregation is enhanced stability against phase separation. Competitive solutes with correlated spectra provide entropic stabilization, whereas cooperative solutes with anti-correlated spectra drive energetic stabilization. Crucially, the maximum stabilizing synergy emerges when the solute spectra are entirely uncorrelated. Furthermore, we show that a nanocrystalline phase can achieve absolute thermodynamic stability only above a certain critical temperature. At lower temperatures, however, cosegregation enhances metastability by elevating the activation energy barrier for second-phase nucleation.
\par
Another vital application of multicomponent segregation relies on the enhanced resistance to grain growth. Cosegregation of uncorrelated or anti-correlated solutes attenuates temperature-induced desegregation, an effect that is most pronounced for anti-correlated species. Furthermore, while the cosegregation of cooperative solutes can suppress grain growth where individual solutes fail, the resulting state remains highly susceptible to the desegregation due to the relatively weak segregation of each distinct solute. 
\par
The implications of multicomponent segregation were demonstrated using polycrystalline silver as a model system. Cosegregation significantly expands the composition space where nanocrystalline alloys remain stable against grain growth, and particularly against phase separation. Increasing the number of cosegregating species systematically lowers the critical temperature of absolute stabilization. However, the most promising solute pairs and triplets include elements that lacked atomic configurations representing solute–GB interactions in the potential's training dataset. Consequently, the calculation accuracy for these specific systems remains limited, necessitating further potential fine-tuning with \textit{ab initio} calculations. Furthermore, the model predicts high GB solute concentrations under these optimal conditions, indicating that a precise descriptive model requires accounting for solute–solute interactions. Nonetheless, this framework enables the transition from single-solute optimization to multicomponent alloy engineering, offering a predictive pathway for synthesizing high-temperature stable nanocrystalline materials.

\section*{Declaration of Competing Interest}
The authors declare that they have no known competing financial interests or personal relationships that could have appeared to influence the work reported in this paper.
\section*{Acknowledgements}
This study represents a contribution made by the Russian Federation to the ITER project (Rosatom contract No. H. 4a.241.09.26.1039) and was supported by the Ministry of Science and Higher Education of the Russian Federation within state assignment  (FFUG–2024– 0034).

\appendix

\section{Gibbs Free Energy in the Spectral Multicomponent Segregation Model}\label{app:gibbs}
In this section we derive the Gibbs energy of segregation for a multicomponent system. For brevity, we consider a ternary system with a matrix $A$ and solutes $B$ and $C$. The generalization to multicomponent systems is straightforward and will be provided further. The configurational energy of segregation $W$ is

\begin{equation}
W = \sum_i \left( N^B_i E^B_i + N^C_i E^C_i \right)\,,
\end{equation}

where $E^{\alpha}_i$ and $N^{\alpha}_i$ are the energy of segregation and the number of sites of type $i$ occupied by solute of type ${\alpha}$. Denote by $F_i$ the number of sites of type $i$. For polycrystals, it is $F_i = f_{gb}\rho_i + (1-f_{gb})\delta_{i,c}$, where $f_{gb}$ is a site fraction of GB atoms and $\rho_i$ is a probability density of GB site type $i$. Subscript $i=c$ denotes the grain interior site with $E_c^{\alpha}=0$. Then the partition function $Z$ is

\begin{equation}
Z = g (N^B_i, N^C_i)\exp\left[ -\frac{W(N^B_i, N^C_i)}{kT}\right]\,,
\end{equation}

where $g$ is the number of distinguishable microstates corresponding to energy $W$:

\begin{equation}
g (N^B_i, N^C_i) = \prod_i \binom{F_i}{N^B_i + N^C_i}\binom{N^B_i + N^C_i}{N^B_i} = \prod_i \frac{F_i !}{N^B_i! N^C_i ! (F_i - N^B_i-N^C_i)! }\,.
\end{equation}

Given the Gibbs energy of segregation $G = -kT\ln Z = W - kT\ln g$, we can use Stirling's approximation to obtain:

\begin{equation}
G  = \sum_i \left[ N^B_i E^B_i + N^C_i E^C_i - kT\left( \ln  F_i - (F_i-N^B_i-N^C_i)\ln (F_i-N^B_i-N^C_i) - N^B_i \ln N^B_i - N^C_i \ln N^C_i\right)\right]\,.
\end{equation}

Finally, it can be written as

\begin{equation}\label{eqA:gibbs}
G = \sum_i F_i\left[ X^B_i E^B_i + X^C_i E^C_i +kT\left( (1-X^B_i-X^C_i)\ln (1-X^B_i-X^C_i) + X^B_i \ln X^B_i + X^C_i \ln X^C_i\right)\right]\,,
\end{equation}

where we define site-specific solute concentration $X^{\alpha}_i = N^{\alpha}_i/F_i$. Since $(1-X^B_i-X^C_i)$ is a concentration of the matrix atoms ($X^A_i$), extension of this equation to multicomponent segregation with $m$ solutes is straightforward:

\begin{equation}\label{eqA:gibbs_m}
G = \sum_i F_i\sum_{\alpha}\left[ X^{\alpha}_i E^{\alpha}_i +kT X^{\alpha}_i\ln (X^{\alpha}_i)\right]\,.
\end{equation}

\section{Derivation of the Nanocrystalline Stability Equation}\label{app:eq1}
This section presents the derivation of the equation for the equilibrium value of $f_{{gb}}$.  The temperature $T$ and total solute concentrations $X^{\alpha}_{tot}$ determine the nanocrystalline state. At equilibrium, the necessary minimum condition yields:

\begin{equation}\label{eqA:eq0}
0 = \derivativec{G_{nc}}{f_{{gb}}}{T, X^{\alpha}_{tot}} = \kgam + G_{{gb}} - G_{c} + f_{{gb}}\derivativec{G_{{gb}}}{f_{{gb}}}{T, X^{\alpha}_{{tot}}} + (1-f_{{gb}})\derivativec{G_{c}}{f_{{gb}}}{T, X^{\alpha}_{{tot}}}\,,
\end{equation}

where we introduced $G_{gb} = \sum_i\rho_i\sum_{\alpha}\left[ E_i^{\alpha}X_i^{\alpha}+kT X_i^{\alpha}\ln X_i^{\alpha}  \right]$ and $G_c = kT\sum_{\alpha}X_c^{\alpha}\ln X_c^{\alpha}$ (here, the sums over $\alpha$ include the matrix $A$). Derivatives in Eq.~\eqref{eqA:eq0} may be rewritten as follows:

\begin{equation}\label{eqA:dggb}
 \derivativec{G_{{gb}}}{f_{{gb}}}{T} = \sum_{\alpha}\sum_i\rho_i \derivative{X_i^{\alpha}}{f_{{gb}}} \left(E_i^{\alpha} + kT\ln\left(\frac{X_i^{\alpha}}{1-X_i^{\alpha}}\right) \right) = \sum_{\alpha}\mu^{\alpha} \derivative{}{f_{{gb}}} \sum_i \rho_i X_i^{\alpha} = \sum_{\alpha}\mu^{\alpha} \derivative{X^{\alpha}_{{gb}}}{f_{{gb}}}\,,
\end{equation}

where $X^{\alpha}_{{gb}}$ is the average GB $\alpha$ solute concentration. Similarly, we can write for $G_{c}$:

\begin{equation}\label{eqA:dgc}
 \derivativec{G_{c}}{f_{{gb}}}{T} = \sum_{\alpha}\mu^{\alpha} \derivative{X^{\alpha}_{c}}{f_{{gb}}}\,.
\end{equation}

The resulting sum $f_{{gb}}\sum_{\alpha}\derivativec{X^{\alpha}_{{gb}}}{f_{{gb}}}{T, X^{\beta}_{{tot}}} + (1-f_{{gb}})\sum_{\alpha}\derivativec{X^{\alpha}_{c}}{f_{{gb}}}{T, X^{\beta}_{{tot}}}$ may be found through differentiation of Eq.~\eqref{eq:X_balance}:

\begin{equation}
 \derivativec{}{f_{{gb}}}{T, X^{\beta}_{{tot}}}\left\{ f_{{gb}}X^{\alpha}_{{gb}} + (1-f_{{gb}})X^{\alpha}_{c} \right\} = \derivativec{X^{\alpha}_{{tot}}}{f_{{gb}}}{T, X^{\beta}_{{tot}}} = 0\,,
\end{equation}

which leads to  

\begin{equation}\label{eqA:dxdf1}
f_{{gb}}\derivativec{X^{\alpha}_{{gb}}}{f_{{gb}}}{T, X^{\beta}_{{tot}}} + (1-f_{{gb}})\derivativec{X^{\alpha}_{c}}{f_{{gb}}}{T, X^{\beta}_{{tot}}} = -(X^{\alpha}_{{gb}} - X^{\alpha}_{c})\,.
\end{equation}

Substituting Eq.~\eqref{eqA:dxdf1} into Eq.~\eqref{eqA:eq0}, we obtain

\begin{equation}\label{eqA:eqGG}
\kgam + G_{{gb}} - G_{c} - \sum_{\alpha}\mu^{\alpha}(X^{\alpha}_{{gb}}- X^{\alpha}_{c}) = 0\,.
\end{equation}

After substituting $\mu^{\alpha}$ from Eq.~\eqref{eq:lagrange} into Eq.~\eqref{eqA:eqGG} and performing some simplifications, we finally obtain:

\begin{equation}
\sum_i \rho_i\ln (1-\sum_{\alpha}X_i^{\alpha}) - \ln(1-\sum_{\alpha}X^{\alpha}_{{c}}) +\frac{\kgam}{kT} = 0\,.
\end{equation}

\def\bibsection{\section*{\refname}}


\begin{thebibliography}{37}%
\makeatletter
\providecommand \@ifxundefined [1]{%
 \@ifx{#1\undefined}
}%
\providecommand \@ifnum [1]{%
 \ifnum #1\expandafter \@firstoftwo
 \else \expandafter \@secondoftwo
 \fi
}%
\providecommand \@ifx [1]{%
 \ifx #1\expandafter \@firstoftwo
 \else \expandafter \@secondoftwo
 \fi
}%
\providecommand \natexlab [1]{#1}%
\providecommand \enquote  [1]{``#1''}%
\providecommand \bibnamefont  [1]{#1}%
\providecommand \bibfnamefont [1]{#1}%
\providecommand \citenamefont [1]{#1}%
\providecommand \href@noop [0]{\@secondoftwo}%
\providecommand \href [0]{\begingroup \@sanitize@url \@href}%
\providecommand \@href[1]{\@@startlink{#1}\@@href}%
\providecommand \@@href[1]{\endgroup#1\@@endlink}%
\providecommand \@sanitize@url [0]{\catcode `\\12\catcode `\$12\catcode `\&12\catcode `\#12\catcode `\^12\catcode `\_12\catcode `\%12\relax}%
\providecommand \@@startlink[1]{}%
\providecommand \@@endlink[0]{}%
\providecommand \url  [0]{\begingroup\@sanitize@url \@url }%
\providecommand \@url [1]{\endgroup\@href {#1}{\urlprefix }}%
\providecommand \urlprefix  [0]{URL }%
\providecommand \Eprint [0]{\href }%
\providecommand \doibase [0]{https://doi.org/}%
\providecommand \selectlanguage [0]{\@gobble}%
\providecommand \bibinfo  [0]{\@secondoftwo}%
\providecommand \bibfield  [0]{\@secondoftwo}%
\providecommand \translation [1]{[#1]}%
\providecommand \BibitemOpen [0]{}%
\providecommand \bibitemStop [0]{}%
\providecommand \bibitemNoStop [0]{.\EOS\space}%
\providecommand \EOS [0]{\spacefactor3000\relax}%
\providecommand \BibitemShut  [1]{\csname bibitem#1\endcsname}%
\let\auto@bib@innerbib\@empty
\bibitem [{\citenamefont {Gleiter}(1989)}]{gleiter}%
  \BibitemOpen
  \bibfield  {author} {\bibinfo {author} {\bibfnamefont {H.}~\bibnamefont {Gleiter}},\ }\bibfield  {title} {\bibinfo {title} {Nanocrystalline materials},\ }\href {https://doi.org/10.1016/0079-6425(89)90001-7} {\bibfield  {journal} {\bibinfo  {journal} {Progress in Materials Science}\ }\textbf {\bibinfo {volume} {33}},\ \bibinfo {pages} {223} (\bibinfo {year} {1989})}\BibitemShut {NoStop}%
\bibitem [{\citenamefont {Petrov}\ \emph {et~al.}(2003)\citenamefont {Petrov}, \citenamefont {Barna}, \citenamefont {Hultman},\ and\ \citenamefont {Greene}}]{petrov2003}%
  \BibitemOpen
  \bibfield  {author} {\bibinfo {author} {\bibfnamefont {I.}~\bibnamefont {Petrov}}, \bibinfo {author} {\bibfnamefont {P.~B.}\ \bibnamefont {Barna}}, \bibinfo {author} {\bibfnamefont {L.}~\bibnamefont {Hultman}},\ and\ \bibinfo {author} {\bibfnamefont {J.~E.}\ \bibnamefont {Greene}},\ }\bibfield  {title} {\bibinfo {title} {Microstructural evolution during film growth},\ }\href {https://doi.org/10.1116/1.1601610} {\bibfield  {journal} {\bibinfo  {journal} {Journal of Vacuum Science \& Technology A}\ }\textbf {\bibinfo {volume} {21}},\ \bibinfo {pages} {S117} (\bibinfo {year} {2003})}\BibitemShut {NoStop}%
\bibitem [{\citenamefont {Peng}\ \emph {et~al.}(2017)\citenamefont {Peng}, \citenamefont {Gong}, \citenamefont {Chen},\ and\ \citenamefont {Liu}}]{peng_thermal_2017}%
  \BibitemOpen
  \bibfield  {author} {\bibinfo {author} {\bibfnamefont {H.~R.}\ \bibnamefont {Peng}}, \bibinfo {author} {\bibfnamefont {M.~M.}\ \bibnamefont {Gong}}, \bibinfo {author} {\bibfnamefont {Y.~Z.}\ \bibnamefont {Chen}},\ and\ \bibinfo {author} {\bibfnamefont {F.}~\bibnamefont {Liu}},\ }\bibfield  {title} {\bibinfo {title} {Thermal stability of nanocrystalline materials: thermodynamics and kinetics},\ }\href {https://doi.org/10.1080/09506608.2016.1257536} {\bibfield  {journal} {\bibinfo  {journal} {International Materials Reviews}\ }\textbf {\bibinfo {volume} {62}},\ \bibinfo {pages} {303} (\bibinfo {year} {2017})}\BibitemShut {NoStop}%
\bibitem [{\citenamefont {Chookajorn}\ \emph {et~al.}(2012)\citenamefont {Chookajorn}, \citenamefont {Murdoch},\ and\ \citenamefont {Schuh}}]{chookajorn_design_2012}%
  \BibitemOpen
  \bibfield  {author} {\bibinfo {author} {\bibfnamefont {T.}~\bibnamefont {Chookajorn}}, \bibinfo {author} {\bibfnamefont {H.~A.}\ \bibnamefont {Murdoch}},\ and\ \bibinfo {author} {\bibfnamefont {C.~A.}\ \bibnamefont {Schuh}},\ }\bibfield  {title} {\bibinfo {title} {Design of stable nanocrystalline alloys},\ }\href {https://doi.org/10.1126/science.1224737} {\bibfield  {journal} {\bibinfo  {journal} {Science}\ }\textbf {\bibinfo {volume} {337}},\ \bibinfo {pages} {951} (\bibinfo {year} {2012})}\BibitemShut {NoStop}%
\bibitem [{\citenamefont {Antonaia}\ \emph {et~al.}(2014)\citenamefont {Antonaia}, \citenamefont {Addonizio}, \citenamefont {Esposito}, \citenamefont {Ferrara}, \citenamefont {Castaldo}, \citenamefont {Guglielmo},\ and\ \citenamefont {D'Angelo}}]{antonaia_adhesion_2014}%
  \BibitemOpen
  \bibfield  {author} {\bibinfo {author} {\bibfnamefont {A.}~\bibnamefont {Antonaia}}, \bibinfo {author} {\bibfnamefont {M.~L.}\ \bibnamefont {Addonizio}}, \bibinfo {author} {\bibfnamefont {S.}~\bibnamefont {Esposito}}, \bibinfo {author} {\bibfnamefont {M.}~\bibnamefont {Ferrara}}, \bibinfo {author} {\bibfnamefont {A.}~\bibnamefont {Castaldo}}, \bibinfo {author} {\bibfnamefont {A.}~\bibnamefont {Guglielmo}},\ and\ \bibinfo {author} {\bibfnamefont {A.}~\bibnamefont {D'Angelo}},\ }\bibfield  {title} {\bibinfo {title} {Adhesion and structural stability enhancement for ag layers deposited on steel in selective solar coatings technology},\ }\href {https://doi.org/10.1016/j.surfcoat.2014.02.045} {\bibfield  {journal} {\bibinfo  {journal} {Surface and Coatings Technology}\ }\textbf {\bibinfo {volume} {255}},\ \bibinfo {pages} {96} (\bibinfo {year} {2014})}\BibitemShut {NoStop}%
\bibitem [{\citenamefont {{Gledhill}}\ \emph {et~al.}(2019)\citenamefont {{Gledhill}}, \citenamefont {{Steyer}}, \citenamefont {{Weiss}},\ and\ \citenamefont {{Hildebrandt}}}]{gledhill_hipims_2019}%
  \BibitemOpen
  \bibfield  {author} {\bibinfo {author} {\bibnamefont {{Gledhill}}}, \bibinfo {author} {\bibnamefont {{Steyer}}}, \bibinfo {author} {\bibnamefont {{Weiss}}},\ and\ \bibinfo {author} {\bibnamefont {{Hildebrandt}}},\ }\bibfield  {title} {\bibinfo {title} {{HiPIMS} and {DC} magnetron sputter-coated silver films for high-temperature durable reflectors},\ }\href {https://doi.org/10.3390/coatings9100593} {\bibfield  {journal} {\bibinfo  {journal} {Coatings}\ }\textbf {\bibinfo {volume} {9}},\ \bibinfo {pages} {593} (\bibinfo {year} {2019})}\BibitemShut {NoStop}%
\bibitem [{\citenamefont {Risse}\ \emph {et~al.}(2008)\citenamefont {Risse}, \citenamefont {Gebhardt}, \citenamefont {Damm}, \citenamefont {Peschel}, \citenamefont {Stöckl}, \citenamefont {Feigl}, \citenamefont {Kirschstein}, \citenamefont {Eberhardt}, \citenamefont {Kaiser},\ and\ \citenamefont {Tünnermann}}]{risse_novel_2008}%
  \BibitemOpen
  \bibfield  {author} {\bibinfo {author} {\bibfnamefont {S.}~\bibnamefont {Risse}}, \bibinfo {author} {\bibfnamefont {A.}~\bibnamefont {Gebhardt}}, \bibinfo {author} {\bibfnamefont {C.}~\bibnamefont {Damm}}, \bibinfo {author} {\bibfnamefont {T.}~\bibnamefont {Peschel}}, \bibinfo {author} {\bibfnamefont {W.}~\bibnamefont {Stöckl}}, \bibinfo {author} {\bibfnamefont {T.}~\bibnamefont {Feigl}}, \bibinfo {author} {\bibfnamefont {S.}~\bibnamefont {Kirschstein}}, \bibinfo {author} {\bibfnamefont {R.}~\bibnamefont {Eberhardt}}, \bibinfo {author} {\bibfnamefont {N.}~\bibnamefont {Kaiser}},\ and\ \bibinfo {author} {\bibfnamefont {A.}~\bibnamefont {Tünnermann}},\ }\bibfield  {title} {\bibinfo {title} {Novel {TMA} telescope based on ultra precise metal mirrors},\ }in\ \href {https://doi.org/10.1117/12.789824} {\emph {\bibinfo {booktitle} {Astronomical Telescopes and Instrumentation}}},\ \bibinfo {editor} {edited by\ \bibinfo {editor} {\bibfnamefont {J.~M.}\ \bibnamefont {Oschmann}, \bibfnamefont {Jr.}}, \bibinfo
  {editor} {\bibfnamefont {M.~W.~M.}\ \bibnamefont {de~Graauw}},\ and\ \bibinfo {editor} {\bibfnamefont {H.~A.}\ \bibnamefont {{MacEwen}}}}\ (\bibinfo {year} {2008})\ p.\ \bibinfo {pages} {701016}\BibitemShut {NoStop}%
\bibitem [{\citenamefont {Sheikh}\ \emph {et~al.}(2008)\citenamefont {Sheikh}, \citenamefont {Connell},\ and\ \citenamefont {Dummer}}]{sheikh_durable_2008}%
  \BibitemOpen
  \bibfield  {author} {\bibinfo {author} {\bibfnamefont {D.~A.}\ \bibnamefont {Sheikh}}, \bibinfo {author} {\bibfnamefont {S.~J.}\ \bibnamefont {Connell}},\ and\ \bibinfo {author} {\bibfnamefont {R.~S.}\ \bibnamefont {Dummer}},\ }\bibfield  {title} {\bibinfo {title} {Durable silver coating for kepler space telescope primary mirror},\ }in\ \href {https://doi.org/10.1117/12.789996} {\emph {\bibinfo {booktitle} {Space Telescopes and Instrumentation 2008: Optical, Infrared, and Millimeter}}},\ Vol.\ \bibinfo {volume} {7010}\ (\bibinfo {year} {2008})\ p.\ \bibinfo {pages} {70104E}\BibitemShut {NoStop}%
\bibitem [{\citenamefont {Samsonov}\ \emph {et~al.}(2022)\citenamefont {Samsonov}, \citenamefont {Tereschenko}, \citenamefont {Mukhin}, \citenamefont {Gubal}, \citenamefont {Kapustin}, \citenamefont {Filimonov}, \citenamefont {Babinov}, \citenamefont {Dmitriev}, \citenamefont {Nikolaev}, \citenamefont {Komarevtsev}, \citenamefont {Koval}, \citenamefont {Litvinov}, \citenamefont {Marchii}, \citenamefont {Razdobarin}, \citenamefont {Snigirev}, \citenamefont {Tolstyakov}, \citenamefont {Marinin}, \citenamefont {Terentev}, \citenamefont {Gorodetsky}, \citenamefont {Zalavutdinov}, \citenamefont {Markin}, \citenamefont {Bukhovets}, \citenamefont {Arkhipushkin}, \citenamefont {Borisov}, \citenamefont {Khripunov}, \citenamefont {Mikhailovskii}, \citenamefont {Modestov}, \citenamefont {Kirienko}, \citenamefont {Buslakov}, \citenamefont {Chernakov}, \citenamefont {Mokeev}, \citenamefont {Kempenaars}, \citenamefont {Shigin},\ and\ \citenamefont {Drapiko}}]{samsonov_large-scale_2022}%
  \BibitemOpen
  \bibfield  {author} {\bibinfo {author} {\bibfnamefont {D.~S.}\ \bibnamefont {Samsonov}}, \bibinfo {author} {\bibfnamefont {I.}~\bibnamefont {Tereschenko}}, \bibinfo {author} {\bibfnamefont {E.~E.}\ \bibnamefont {Mukhin}}, \bibinfo {author} {\bibfnamefont {A.}~\bibnamefont {Gubal}}, \bibinfo {author} {\bibfnamefont {Y.}~\bibnamefont {Kapustin}}, \bibinfo {author} {\bibfnamefont {V.}~\bibnamefont {Filimonov}}, \bibinfo {author} {\bibfnamefont {N.~A.}\ \bibnamefont {Babinov}}, \bibinfo {author} {\bibfnamefont {A.~M.}\ \bibnamefont {Dmitriev}}, \bibinfo {author} {\bibfnamefont {A.}~\bibnamefont {Nikolaev}}, \bibinfo {author} {\bibfnamefont {I.}~\bibnamefont {Komarevtsev}}, \bibinfo {author} {\bibfnamefont {A.}~\bibnamefont {Koval}}, \bibinfo {author} {\bibfnamefont {A.~E.}\ \bibnamefont {Litvinov}}, \bibinfo {author} {\bibfnamefont {G.}~\bibnamefont {Marchii}}, \bibinfo {author} {\bibfnamefont {A.}~\bibnamefont {Razdobarin}}, \bibinfo {author} {\bibfnamefont {L.}~\bibnamefont {Snigirev}}, \bibinfo {author}
  {\bibfnamefont {S.}~\bibnamefont {Tolstyakov}}, \bibinfo {author} {\bibfnamefont {G.}~\bibnamefont {Marinin}}, \bibinfo {author} {\bibfnamefont {D.}~\bibnamefont {Terentev}}, \bibinfo {author} {\bibfnamefont {A.~E.}\ \bibnamefont {Gorodetsky}}, \bibinfo {author} {\bibfnamefont {R.~K.}\ \bibnamefont {Zalavutdinov}}, \bibinfo {author} {\bibfnamefont {A.~V.}\ \bibnamefont {Markin}}, \bibinfo {author} {\bibfnamefont {V.}~\bibnamefont {Bukhovets}}, \bibinfo {author} {\bibfnamefont {I.}~\bibnamefont {Arkhipushkin}}, \bibinfo {author} {\bibfnamefont {A.}~\bibnamefont {Borisov}}, \bibinfo {author} {\bibfnamefont {V.~I.}\ \bibnamefont {Khripunov}}, \bibinfo {author} {\bibfnamefont {V.}~\bibnamefont {Mikhailovskii}}, \bibinfo {author} {\bibfnamefont {V.}~\bibnamefont {Modestov}}, \bibinfo {author} {\bibfnamefont {I.}~\bibnamefont {Kirienko}}, \bibinfo {author} {\bibfnamefont {I.}~\bibnamefont {Buslakov}}, \bibinfo {author} {\bibfnamefont {P.~V.}\ \bibnamefont {Chernakov}}, \bibinfo {author} {\bibfnamefont
  {A.}~\bibnamefont {Mokeev}}, \bibinfo {author} {\bibfnamefont {M.}~\bibnamefont {Kempenaars}}, \bibinfo {author} {\bibfnamefont {P.~A.}\ \bibnamefont {Shigin}},\ and\ \bibinfo {author} {\bibfnamefont {E.}~\bibnamefont {Drapiko}},\ }\bibfield  {title} {\bibinfo {title} {Large-scale collecting mirrors for {ITER} optical diagnostic},\ }\bibfield  {journal} {\bibinfo  {journal} {Nuclear Fusion}\ }\href {https://doi.org/10.1088/1741-4326/ac544d} {10.1088/1741-4326/ac544d} (\bibinfo {year} {2022})\BibitemShut {NoStop}%
\bibitem [{\citenamefont {Jacquet}\ \emph {et~al.}(2016)\citenamefont {Jacquet}, \citenamefont {Podor}, \citenamefont {Ravaux}, \citenamefont {Teisseire}, \citenamefont {Gozhyk}, \citenamefont {Jupille},\ and\ \citenamefont {Lazzari}}]{jacquet_grain_2016}%
  \BibitemOpen
  \bibfield  {author} {\bibinfo {author} {\bibfnamefont {P.}~\bibnamefont {Jacquet}}, \bibinfo {author} {\bibfnamefont {R.}~\bibnamefont {Podor}}, \bibinfo {author} {\bibfnamefont {J.}~\bibnamefont {Ravaux}}, \bibinfo {author} {\bibfnamefont {J.}~\bibnamefont {Teisseire}}, \bibinfo {author} {\bibfnamefont {I.}~\bibnamefont {Gozhyk}}, \bibinfo {author} {\bibfnamefont {J.}~\bibnamefont {Jupille}},\ and\ \bibinfo {author} {\bibfnamefont {R.}~\bibnamefont {Lazzari}},\ }\bibfield  {title} {\bibinfo {title} {Grain growth: The key to understand solid-state dewetting of silver thin films},\ }\href {https://doi.org/10.1016/j.scriptamat.2016.01.005} {\bibfield  {journal} {\bibinfo  {journal} {Scripta Materialia}\ }\textbf {\bibinfo {volume} {115}},\ \bibinfo {pages} {128} (\bibinfo {year} {2016})}\BibitemShut {NoStop}%
\bibitem [{\citenamefont {Chason}\ and\ \citenamefont {Guduru}(2016)}]{chason_tutorial_2016}%
  \BibitemOpen
  \bibfield  {author} {\bibinfo {author} {\bibfnamefont {E.}~\bibnamefont {Chason}}\ and\ \bibinfo {author} {\bibfnamefont {P.~R.}\ \bibnamefont {Guduru}},\ }\bibfield  {title} {\bibinfo {title} {Tutorial: Understanding residual stress in polycrystalline thin films through real-time measurements and physical models},\ }\href {https://doi.org/10.1063/1.4949263} {\bibfield  {journal} {\bibinfo  {journal} {Journal of Applied Physics}\ }\textbf {\bibinfo {volume} {119}},\ \bibinfo {pages} {191101} (\bibinfo {year} {2016})}\BibitemShut {NoStop}%
\bibitem [{\citenamefont {Tereshchenko}\ \emph {et~al.}(2025)\citenamefont {Tereshchenko}, \citenamefont {Marchiy}, \citenamefont {Samsonov}, \citenamefont {Mukhin}, \citenamefont {Kapustin}, \citenamefont {Kalganov}, \citenamefont {Gubal},\ and\ \citenamefont {Komarevtcev}}]{tereshenko2025}%
  \BibitemOpen
  \bibfield  {author} {\bibinfo {author} {\bibfnamefont {I.~B.}\ \bibnamefont {Tereshchenko}}, \bibinfo {author} {\bibfnamefont {G.~V.}\ \bibnamefont {Marchiy}}, \bibinfo {author} {\bibfnamefont {D.~S.}\ \bibnamefont {Samsonov}}, \bibinfo {author} {\bibfnamefont {E.~E.}\ \bibnamefont {Mukhin}}, \bibinfo {author} {\bibfnamefont {Y.~V.}\ \bibnamefont {Kapustin}}, \bibinfo {author} {\bibfnamefont {V.~D.}\ \bibnamefont {Kalganov}}, \bibinfo {author} {\bibfnamefont {A.~R.}\ \bibnamefont {Gubal}},\ and\ \bibinfo {author} {\bibfnamefont {I.~M.}\ \bibnamefont {Komarevtcev}},\ }\bibfield  {title} {\bibinfo {title} {Highly reflective silver mirror under annealing and hydrothermal exposure},\ }\href {https://doi.org/10.61011/TPL.2025.03.60712.20136} {\bibfield  {journal} {\bibinfo  {journal} {Technical Physics Letters}\ }\textbf {\bibinfo {volume} {51}},\ \bibinfo {pages} {22} (\bibinfo {year} {2025})}\BibitemShut {NoStop}%
\bibitem [{\citenamefont {Weissm{\"u}ller}(1993)}]{weissmuller_alloy_1993}%
  \BibitemOpen
  \bibfield  {author} {\bibinfo {author} {\bibfnamefont {J.}~\bibnamefont {Weissm{\"u}ller}},\ }\bibfield  {title} {\bibinfo {title} {Alloy effects in nanostructures},\ }\href {https://doi.org/10.1016/0965-9773(93)90088-S} {\bibfield  {journal} {\bibinfo  {journal} {Nanostructured Materials}\ }\bibinfo {series} {Proceedings of the First International Conference on Nanostructured Materials},\ \textbf {\bibinfo {volume} {3}},\ \bibinfo {pages} {261} (\bibinfo {year} {1993})}\BibitemShut {NoStop}%
\bibitem [{\citenamefont {Trelewicz}\ and\ \citenamefont {Schuh}(2009)}]{trelewicz_grain_2009}%
  \BibitemOpen
  \bibfield  {author} {\bibinfo {author} {\bibfnamefont {J.~R.}\ \bibnamefont {Trelewicz}}\ and\ \bibinfo {author} {\bibfnamefont {C.~A.}\ \bibnamefont {Schuh}},\ }\bibfield  {title} {\bibinfo {title} {Grain boundary segregation and thermodynamically stable binary nanocrystalline alloys},\ }\href {https://doi.org/10.1103/PhysRevB.79.094112} {\bibfield  {journal} {\bibinfo  {journal} {Physical Review B}\ }\textbf {\bibinfo {volume} {79}},\ \bibinfo {pages} {094112} (\bibinfo {year} {2009})}\BibitemShut {NoStop}%
\bibitem [{\citenamefont {Matson}\ \emph {et~al.}(2026)\citenamefont {Matson}, \citenamefont {Tuchinda},\ and\ \citenamefont {Schuh}}]{matson_overview_2026}%
  \BibitemOpen
  \bibfield  {author} {\bibinfo {author} {\bibfnamefont {T.~P.}\ \bibnamefont {Matson}}, \bibinfo {author} {\bibfnamefont {N.}~\bibnamefont {Tuchinda}},\ and\ \bibinfo {author} {\bibfnamefont {C.~A.}\ \bibnamefont {Schuh}},\ }\bibfield  {title} {\bibinfo {title} {Overview: The spectral model of grain boundary segregation},\ }\href {https://doi.org/10.1016/j.actamat.2026.122109} {\bibfield  {journal} {\bibinfo  {journal} {Acta Materialia}\ }\textbf {\bibinfo {volume} {313}},\ \bibinfo {pages} {122109} (\bibinfo {year} {2026})}\BibitemShut {NoStop}%
\bibitem [{\citenamefont {Wagih}\ \emph {et~al.}(2025)\citenamefont {Wagih}, \citenamefont {Naunheim}, \citenamefont {Lei},\ and\ \citenamefont {Schuh}}]{wagih_designing_2025}%
  \BibitemOpen
  \bibfield  {author} {\bibinfo {author} {\bibfnamefont {M.}~\bibnamefont {Wagih}}, \bibinfo {author} {\bibfnamefont {Y.}~\bibnamefont {Naunheim}}, \bibinfo {author} {\bibfnamefont {T.}~\bibnamefont {Lei}},\ and\ \bibinfo {author} {\bibfnamefont {C.~A.}\ \bibnamefont {Schuh}},\ }\bibfield  {title} {\bibinfo {title} {Designing for cooperative grain boundary segregation in multicomponent alloys},\ }\href {https://doi.org/10.1073/pnas.2511930122} {\bibfield  {journal} {\bibinfo  {journal} {Proceedings of the National Academy of Sciences}\ }\textbf {\bibinfo {volume} {122}},\ \bibinfo {pages} {e2511930122} (\bibinfo {year} {2025})}\BibitemShut {NoStop}%
\bibitem [{\citenamefont {Guttmann}(1975)}]{guttmann_1975}%
  \BibitemOpen
  \bibfield  {author} {\bibinfo {author} {\bibfnamefont {M.}~\bibnamefont {Guttmann}},\ }\bibfield  {title} {\bibinfo {title} {Equilibrium segregation in a ternary solution: A model for temper embrittlement},\ }\href {https://doi.org/10.1016/0039-6028(75)90125-9} {\bibfield  {journal} {\bibinfo  {journal} {Surface Science}\ }\textbf {\bibinfo {volume} {53}},\ \bibinfo {pages} {213} (\bibinfo {year} {1975})}\BibitemShut {NoStop}%
\bibitem [{\citenamefont {Guttmann}(1995)}]{guttmann_1995}%
  \BibitemOpen
  \bibfield  {author} {\bibinfo {author} {\bibfnamefont {M.}~\bibnamefont {Guttmann}},\ }\bibfield  {title} {\bibinfo {title} {Thermochemical interactions versus site competition in grain boundary segregation and embrittlement in multicomponent systems},\ }\href {https://doi.org/10.1051/jp4:1995707} {\bibfield  {journal} {\bibinfo  {journal} {Journal de Physique IV (Proceedings)}\ }\textbf {\bibinfo {volume} {05}},\ \bibinfo {pages} {C7} (\bibinfo {year} {1995})}\BibitemShut {NoStop}%
\bibitem [{\citenamefont {Marchiy}\ \emph {et~al.}(2025)\citenamefont {Marchiy}, \citenamefont {Samsonov},\ and\ \citenamefont {Mukhin}}]{marchiy_spectral_2025}%
  \BibitemOpen
  \bibfield  {author} {\bibinfo {author} {\bibfnamefont {G.}~\bibnamefont {Marchiy}}, \bibinfo {author} {\bibfnamefont {D.}~\bibnamefont {Samsonov}},\ and\ \bibinfo {author} {\bibfnamefont {E.}~\bibnamefont {Mukhin}},\ }\bibfield  {title} {\bibinfo {title} {Spectral model for grain boundary segregation in systems with strong solute-solute interactions},\ }\href {https://doi.org/10.1016/j.actamat.2025.121044} {\bibfield  {journal} {\bibinfo  {journal} {Acta Materialia}\ }\textbf {\bibinfo {volume} {294}},\ \bibinfo {pages} {121044} (\bibinfo {year} {2025})}\BibitemShut {NoStop}%
\bibitem [{\citenamefont {Hildebrandt}\ and\ \citenamefont {Glasser}(1994)}]{hildebrandt_predicting_1994}%
  \BibitemOpen
  \bibfield  {author} {\bibinfo {author} {\bibfnamefont {D.}~\bibnamefont {Hildebrandt}}\ and\ \bibinfo {author} {\bibfnamefont {D.}~\bibnamefont {Glasser}},\ }\bibfield  {title} {\bibinfo {title} {Predicting phase and chemical equilibrium using the convex hull of the gibbs free energy},\ }\href {https://doi.org/10.1016/0923-0467(94)00202-9} {\bibfield  {journal} {\bibinfo  {journal} {The Chemical Engineering Journal and the Biochemical Engineering Journal}\ }\textbf {\bibinfo {volume} {54}},\ \bibinfo {pages} {187} (\bibinfo {year} {1994})}\BibitemShut {NoStop}%
\bibitem [{\citenamefont {Wagih}\ and\ \citenamefont {Schuh}(2021)}]{wagih_thermodynamics_2021}%
  \BibitemOpen
  \bibfield  {author} {\bibinfo {author} {\bibfnamefont {M.}~\bibnamefont {Wagih}}\ and\ \bibinfo {author} {\bibfnamefont {C.~A.}\ \bibnamefont {Schuh}},\ }\bibfield  {title} {\bibinfo {title} {Thermodynamics and design of nanocrystalline alloys using grain boundary segregation spectra},\ }\href {https://doi.org/10.1016/j.actamat.2021.117177} {\bibfield  {journal} {\bibinfo  {journal} {Acta Materialia}\ }\textbf {\bibinfo {volume} {217}},\ \bibinfo {pages} {117177} (\bibinfo {year} {2021})}\BibitemShut {NoStop}%
\bibitem [{\citenamefont {Laughlin}\ and\ \citenamefont {Soffa}(2018)}]{laughlin_third_2018}%
  \BibitemOpen
  \bibfield  {author} {\bibinfo {author} {\bibfnamefont {D.~E.}\ \bibnamefont {Laughlin}}\ and\ \bibinfo {author} {\bibfnamefont {W.~A.}\ \bibnamefont {Soffa}},\ }\bibfield  {title} {\bibinfo {title} {The third law of thermodynamics: Phase equilibria and phase diagrams at low temperatures},\ }\href {https://doi.org/10.1016/j.actamat.2017.11.037} {\bibfield  {journal} {\bibinfo  {journal} {Acta Materialia}\ }\textbf {\bibinfo {volume} {145}},\ \bibinfo {pages} {49} (\bibinfo {year} {2018})}\BibitemShut {NoStop}%
\bibitem [{\citenamefont {Fedorov}(2010)}]{fedorov_third_2010}%
  \BibitemOpen
  \bibfield  {author} {\bibinfo {author} {\bibfnamefont {P.~P.}\ \bibnamefont {Fedorov}},\ }\bibfield  {title} {\bibinfo {title} {Third law of thermodynamics as applied to phase diagrams},\ }\href {https://doi.org/10.1134/S0036023610110100} {\bibfield  {journal} {\bibinfo  {journal} {Russian Journal of Inorganic Chemistry}\ }\textbf {\bibinfo {volume} {55}},\ \bibinfo {pages} {1722} (\bibinfo {year} {2010})}\BibitemShut {NoStop}%
\bibitem [{\citenamefont {Porter}\ \emph {et~al.}(2021)\citenamefont {Porter}, \citenamefont {Easterling},\ and\ \citenamefont {Sherif}}]{porter_phase_2021}%
  \BibitemOpen
  \bibfield  {author} {\bibinfo {author} {\bibfnamefont {D.}~\bibnamefont {Porter}}, \bibinfo {author} {\bibfnamefont {K.}~\bibnamefont {Easterling}},\ and\ \bibinfo {author} {\bibfnamefont {M.}~\bibnamefont {Sherif}},\ }\href {https://doi.org/10.1201/9781003011804} {\emph {\bibinfo {title} {Phase Transformations in Metals and Alloys}}}\ (\bibinfo  {publisher} {CRC Press},\ \bibinfo {year} {2021})\BibitemShut {NoStop}%
\bibitem [{\citenamefont {Jiao}\ and\ \citenamefont {Schuh}(2018)}]{jiao_nanocrystalline_2018}%
  \BibitemOpen
  \bibfield  {author} {\bibinfo {author} {\bibfnamefont {Z.~B.}\ \bibnamefont {Jiao}}\ and\ \bibinfo {author} {\bibfnamefont {C.~A.}\ \bibnamefont {Schuh}},\ }\bibfield  {title} {\bibinfo {title} {Nanocrystalline Ag-W alloys lose stability upon solute desegregation from grain boundaries},\ }\href {https://doi.org} {\bibfield  {journal} {\bibinfo {journal} {Acta Materialia}\ }\textbf {\bibinfo {volume} {161}},\ \bibinfo {pages} {194--206} (\bibinfo {year} {2018})}\BibitemShut {NoStop}%
\bibitem [{\citenamefont {Xu}\ \emph {et~al.}(2025)\citenamefont {Xu}, \citenamefont {Bu}, \citenamefont {Pan}, \citenamefont {Lindgren}, \citenamefont {Wu}, \citenamefont {Wang}, \citenamefont {Liu}, \citenamefont {Song}, \citenamefont {Xu}, \citenamefont {Li}, \citenamefont {Hainer}, \citenamefont {Svensson}, \citenamefont {Wiktor}, \citenamefont {Zhao}, \citenamefont {Huang}, \citenamefont {Qian}, \citenamefont {Zhang}, \citenamefont {Zeng}, \citenamefont {Zhang}, \citenamefont {Tang}, \citenamefont {Xiao}, \citenamefont {Yan}, \citenamefont {Shi}, \citenamefont {Liang}, \citenamefont {Wang}, \citenamefont {Liang}, \citenamefont {Cao}, \citenamefont {Wang}, \citenamefont {Ying}, \citenamefont {Xu}, \citenamefont {Chen}, \citenamefont {Zhang}, \citenamefont {Chen}, \citenamefont {Wu}, \citenamefont {Jiang}, \citenamefont {Berger}, \citenamefont {Li}, \citenamefont {Chen}, \citenamefont {Gabourie}, \citenamefont {Dong}, \citenamefont {Xiong}, \citenamefont {Wei}, \citenamefont {Chen}, \citenamefont {Xu},
  \citenamefont {Ding}, \citenamefont {Sun}, \citenamefont {Ala-Nissila}, \citenamefont {Harju}, \citenamefont {Zheng}, \citenamefont {Guan}, \citenamefont {Erhart}, \citenamefont {Sun}, \citenamefont {Ouyang}, \citenamefont {Su},\ and\ \citenamefont {Fan}}]{gpumd}%
  \BibitemOpen
  \bibfield  {author} {\bibinfo {author} {\bibfnamefont {K.}~\bibnamefont {Xu}}, \bibinfo {author} {\bibfnamefont {H.}~\bibnamefont {Bu}}, \bibinfo {author} {\bibfnamefont {S.}~\bibnamefont {Pan}}, \bibinfo {author} {\bibfnamefont {E.}~\bibnamefont {Lindgren}}, \bibinfo {author} {\bibfnamefont {Y.}~\bibnamefont {Wu}}, \bibinfo {author} {\bibfnamefont {Y.}~\bibnamefont {Wang}}, \bibinfo {author} {\bibfnamefont {J.}~\bibnamefont {Liu}}, \bibinfo {author} {\bibfnamefont {K.}~\bibnamefont {Song}}, \bibinfo {author} {\bibfnamefont {B.}~\bibnamefont {Xu}}, \bibinfo {author} {\bibfnamefont {Y.}~\bibnamefont {Li}}, \bibinfo {author} {\bibfnamefont {T.}~\bibnamefont {Hainer}}, \bibinfo {author} {\bibfnamefont {L.}~\bibnamefont {Svensson}}, \bibinfo {author} {\bibfnamefont {J.}~\bibnamefont {Wiktor}}, \bibinfo {author} {\bibfnamefont {R.}~\bibnamefont {Zhao}}, \bibinfo {author} {\bibfnamefont {H.}~\bibnamefont {Huang}}, \bibinfo {author} {\bibfnamefont {C.}~\bibnamefont {Qian}}, \bibinfo {author} {\bibfnamefont
  {S.}~\bibnamefont {Zhang}}, \bibinfo {author} {\bibfnamefont {Z.}~\bibnamefont {Zeng}}, \bibinfo {author} {\bibfnamefont {B.}~\bibnamefont {Zhang}}, \bibinfo {author} {\bibfnamefont {B.}~\bibnamefont {Tang}}, \bibinfo {author} {\bibfnamefont {Y.}~\bibnamefont {Xiao}}, \bibinfo {author} {\bibfnamefont {Z.}~\bibnamefont {Yan}}, \bibinfo {author} {\bibfnamefont {J.}~\bibnamefont {Shi}}, \bibinfo {author} {\bibfnamefont {Z.}~\bibnamefont {Liang}}, \bibinfo {author} {\bibfnamefont {J.}~\bibnamefont {Wang}}, \bibinfo {author} {\bibfnamefont {T.}~\bibnamefont {Liang}}, \bibinfo {author} {\bibfnamefont {S.}~\bibnamefont {Cao}}, \bibinfo {author} {\bibfnamefont {Y.}~\bibnamefont {Wang}}, \bibinfo {author} {\bibfnamefont {P.}~\bibnamefont {Ying}}, \bibinfo {author} {\bibfnamefont {N.}~\bibnamefont {Xu}}, \bibinfo {author} {\bibfnamefont {C.}~\bibnamefont {Chen}}, \bibinfo {author} {\bibfnamefont {Y.}~\bibnamefont {Zhang}}, \bibinfo {author} {\bibfnamefont {Z.}~\bibnamefont {Chen}}, \bibinfo {author} {\bibfnamefont
  {X.}~\bibnamefont {Wu}}, \bibinfo {author} {\bibfnamefont {W.}~\bibnamefont {Jiang}}, \bibinfo {author} {\bibfnamefont {E.}~\bibnamefont {Berger}}, \bibinfo {author} {\bibfnamefont {Y.}~\bibnamefont {Li}}, \bibinfo {author} {\bibfnamefont {S.}~\bibnamefont {Chen}}, \bibinfo {author} {\bibfnamefont {A.~J.}\ \bibnamefont {Gabourie}}, \bibinfo {author} {\bibfnamefont {H.}~\bibnamefont {Dong}}, \bibinfo {author} {\bibfnamefont {S.}~\bibnamefont {Xiong}}, \bibinfo {author} {\bibfnamefont {N.}~\bibnamefont {Wei}}, \bibinfo {author} {\bibfnamefont {Y.}~\bibnamefont {Chen}}, \bibinfo {author} {\bibfnamefont {J.}~\bibnamefont {Xu}}, \bibinfo {author} {\bibfnamefont {F.}~\bibnamefont {Ding}}, \bibinfo {author} {\bibfnamefont {Z.}~\bibnamefont {Sun}}, \bibinfo {author} {\bibfnamefont {T.}~\bibnamefont {Ala-Nissila}}, \bibinfo {author} {\bibfnamefont {A.}~\bibnamefont {Harju}}, \bibinfo {author} {\bibfnamefont {J.}~\bibnamefont {Zheng}}, \bibinfo {author} {\bibfnamefont {P.}~\bibnamefont {Guan}}, \bibinfo {author}
  {\bibfnamefont {P.}~\bibnamefont {Erhart}}, \bibinfo {author} {\bibfnamefont {J.}~\bibnamefont {Sun}}, \bibinfo {author} {\bibfnamefont {W.}~\bibnamefont {Ouyang}}, \bibinfo {author} {\bibfnamefont {Y.}~\bibnamefont {Su}},\ and\ \bibinfo {author} {\bibfnamefont {Z.}~\bibnamefont {Fan}},\ }\bibfield  {title} {\bibinfo {title} {Gpumd 4.0: A high-performance molecular dynamics package for versatile materials simulations with machine-learned potentials},\ }\href {https://doi.org/10.1002/mgea.70028} {\bibfield  {journal} {\bibinfo  {journal} {Materials Genome Engineering Advances}\ }\textbf {\bibinfo {volume} {3}},\ \bibinfo {pages} {e70028} (\bibinfo {year} {2025})}\BibitemShut {NoStop}%
\bibitem [{\citenamefont {Liang}\ \emph {et~al.}(2026)\citenamefont {Liang}, \citenamefont {Xu}, \citenamefont {Lindgren}, \citenamefont {Chen}, \citenamefont {Zhao}, \citenamefont {Liu}, \citenamefont {Berger}, \citenamefont {Tang}, \citenamefont {Zhang}, \citenamefont {Wang}, \citenamefont {Song}, \citenamefont {Ying}, \citenamefont {Xu}, \citenamefont {Dong}, \citenamefont {Chen}, \citenamefont {Erhart}, \citenamefont {Fan}, \citenamefont {Ala-Nissila},\ and\ \citenamefont {Xu}}]{nep89}%
  \BibitemOpen
  \bibfield  {author} {\bibinfo {author} {\bibfnamefont {T.}\ \bibnamefont {Liang}}, \bibinfo {author} {\bibfnamefont {K.}\ \bibnamefont {Xu}}, \bibinfo {author} {\bibfnamefont {E.}\ \bibnamefont {Lindgren}}, \bibinfo {author} {\bibfnamefont {Z.}\ \bibnamefont {Chen}}, \bibinfo {author} {\bibfnamefont {R.}\ \bibnamefont {Zhao}}, \bibinfo {author} {\bibfnamefont {J.}\ \bibnamefont {Liu}}, \bibinfo {author} {\bibfnamefont {E.}\ \bibnamefont {Berger}}, \bibinfo {author} {\bibfnamefont {B.}\ \bibnamefont {Tang}}, \bibinfo {author} {\bibfnamefont {B.}\ \bibnamefont {Zhang}}, \bibinfo {author} {\bibfnamefont {Y.}\ \bibnamefont {Wang}}, \bibinfo {author} {\bibfnamefont {K.}\ \bibnamefont {Song}}, \bibinfo {author} {\bibfnamefont {P.}\ \bibnamefont {Ying}}, \bibinfo {author} {\bibfnamefont {N.}\ \bibnamefont {Xu}}, \bibinfo {author} {\bibfnamefont {H.}\ \bibnamefont {Dong}}, \bibinfo {author} {\bibfnamefont {S.}\ \bibnamefont {Chen}}, \bibinfo {author} {\bibfnamefont {P.}\ \bibnamefont {Erhart}}, \bibinfo {author} {\bibfnamefont {Z.}\ \bibnamefont {Fan}}, \bibinfo {author} {\bibfnamefont {T.}\ \bibnamefont {Ala-Nissila}},\ and\ \bibinfo {author} {\bibfnamefont {J.}\ \bibnamefont {Xu}},\ }\bibfield  {title} {\bibinfo {title} {{NEP}89: universal neuroevolution potential for inorganic and organic materials across 89 elements},\ }\href {https://doi.org} {\bibfield  {journal} {\bibinfo {journal} {Nature Computational Science}\ }\textbf {\bibinfo {volume} {6}},\ \bibinfo {pages} {789--801} (\bibinfo {year} {2026})}\BibitemShut {NoStop}%
\bibitem [{\citenamefont {Stukowski}(2010)}]
{stukowski_visualization_2010}%
  \BibitemOpen
  \bibfield  {author} {\bibinfo {author} {\bibfnamefont {A.}~\bibnamefont {Stukowski}},\ }\bibfield  {title} {\bibinfo {title} {Visualization and analysis of atomistic simulation data with {OVITO}-the open visualization tool},\ }\bibfield  {journal} {\bibinfo  {journal} {Modelling and Simulation in Materials Science and Engineering}\ }\textbf {\bibinfo {volume} {18}},\ \href {https://doi.org/10.1088/0965-0393/18/1/015012} {10.1088/0965-0393/18/1/015012} (\bibinfo {year} {2010})\BibitemShut {NoStop}%
\bibitem [{\citenamefont {Pedregosa}\ \emph {et~al.}(2011)\citenamefont {Pedregosa}, \citenamefont {Varoquaux}, \citenamefont {Gramfort}, \citenamefont {Michel}, \citenamefont {Thirion}, \citenamefont {Grisel}, \citenamefont {Blondel}, \citenamefont {Prettenhofer}, \citenamefont {Weiss}, \citenamefont {Dubourg} \emph {et~al.}}]{scikit}%
  \BibitemOpen
  \bibfield  {author} {\bibinfo {author} {\bibfnamefont {F.}~\bibnamefont {Pedregosa}}, \bibinfo {author} {\bibfnamefont {G.}~\bibnamefont {Varoquaux}}, \bibinfo {author} {\bibfnamefont {A.}~\bibnamefont {Gramfort}}, \bibinfo {author} {\bibfnamefont {V.}~\bibnamefont {Michel}}, \bibinfo {author} {\bibfnamefont {B.}~\bibnamefont {Thirion}}, \bibinfo {author} {\bibfnamefont {O.}~\bibnamefont {Grisel}}, \bibinfo {author} {\bibfnamefont {M.}~\bibnamefont {Blondel}}, \bibinfo {author} {\bibfnamefont {P.}~\bibnamefont {Prettenhofer}}, \bibinfo {author} {\bibfnamefont {R.}~\bibnamefont {Weiss}}, \bibinfo {author} {\bibfnamefont {V.}~\bibnamefont {Dubourg}}, \emph {et~al.},\ }\bibfield  {title} {\bibinfo {title} {Scikit-learn: Machine learning in python},\ }\href@noop {} {\bibfield  {journal} {\bibinfo  {journal} {The Journal of Machine Learning Research}\ }\textbf {\bibinfo {volume} {12}},\ \bibinfo {pages} {2825} (\bibinfo {year} {2011})}\BibitemShut {NoStop}%
\bibitem [{\citenamefont {Hirel}(2015)}]{hirel_atomsk_2015}%
  \BibitemOpen
  \bibfield  {author} {\bibinfo {author} {\bibfnamefont {P.}~\bibnamefont {Hirel}},\ }\bibfield  {title} {\bibinfo {title} {Atomsk: A tool for manipulating and converting atomic data files},\ }\href {https://doi.org/10.1016/j.cpc.2015.07.012} {\bibfield  {journal} {\bibinfo  {journal} {Computer Physics Communications}\ }\textbf {\bibinfo {volume} {197}},\ \bibinfo {pages} {212} (\bibinfo {year} {2015})}\BibitemShut {NoStop}%
\bibitem [{\citenamefont {Bitzek}\ \emph {et~al.}(2006)\citenamefont {Bitzek}, \citenamefont {Koskinen}, \citenamefont {G\"ahler}, \citenamefont {Moseler},\ and\ \citenamefont {Gumbsch}}]{fire}%
  \BibitemOpen
  \bibfield  {author} {\bibinfo {author} {\bibfnamefont {E.}~\bibnamefont {Bitzek}}, \bibinfo {author} {\bibfnamefont {P.}~\bibnamefont {Koskinen}}, \bibinfo {author} {\bibfnamefont {F.}~\bibnamefont {G\"ahler}}, \bibinfo {author} {\bibfnamefont {M.}~\bibnamefont {Moseler}},\ and\ \bibinfo {author} {\bibfnamefont {P.}~\bibnamefont {Gumbsch}},\ }\bibfield  {title} {\bibinfo {title} {Structural relaxation made simple},\ }\href {https://doi.org/10.1103/PhysRevLett.97.170201} {\bibfield  {journal} {\bibinfo  {journal} {Phys. Rev. Lett.}\ }\textbf {\bibinfo {volume} {97}},\ \bibinfo {pages} {170201} (\bibinfo {year} {2006})}\BibitemShut {NoStop}%
\bibitem [{\citenamefont {Berendsen}\ \emph {et~al.}(1984)\citenamefont {Berendsen}, \citenamefont {Postma}, \citenamefont {Van~Gunsteren}, \citenamefont {{DiNola}},\ and\ \citenamefont {Haak}}]{berendsen_molecular_1984}%
  \BibitemOpen
  \bibfield  {author} {\bibinfo {author} {\bibfnamefont {H.~J.}\ \bibnamefont {Berendsen}}, \bibinfo {author} {\bibfnamefont {J.~v.}\ \bibnamefont {Postma}}, \bibinfo {author} {\bibfnamefont {W.~F.}\ \bibnamefont {Van~Gunsteren}}, \bibinfo {author} {\bibfnamefont {A.}~\bibnamefont {{DiNola}}},\ and\ \bibinfo {author} {\bibfnamefont {J.~R.}\ \bibnamefont {Haak}},\ }\bibfield  {title} {\bibinfo {title} {Molecular dynamics with coupling to an external bath},\ }\href@noop {} {\bibfield  {journal} {\bibinfo  {journal} {The Journal of Chemical Physics}\ }\textbf {\bibinfo {volume} {81}},\ \bibinfo {pages} {3684} (\bibinfo {year} {1984})}\BibitemShut {NoStop}%
\bibitem [{\citenamefont {Wagih}\ \emph {et~al.}(2020)\citenamefont {Wagih}, \citenamefont {Larsen},\ and\ \citenamefont {Schuh}}]{wagih_learning_2020}%
  \BibitemOpen
  \bibfield  {author} {\bibinfo {author} {\bibfnamefont {M.}~\bibnamefont {Wagih}}, \bibinfo {author} {\bibfnamefont {P.~M.}\ \bibnamefont {Larsen}},\ and\ \bibinfo {author} {\bibfnamefont {C.~A.}\ \bibnamefont {Schuh}},\ }\bibfield  {title} {\bibinfo {title} {Learning grain boundary segregation energy spectra in polycrystals},\ }\href {https://doi.org/10.1038/s41467-020-20083-6} {\bibfield  {journal} {\bibinfo  {journal} {Nature Communications}\ }\textbf {\bibinfo {volume} {11}},\ \bibinfo {pages} {6376} (\bibinfo {year} {2020})}\BibitemShut {NoStop}%
\bibitem [{\citenamefont {Bartók}\ \emph {et~al.}(2013)\citenamefont {Bartók}, \citenamefont {Kondor},\ and\ \citenamefont {Csányi}}]{bartok_representing_2013}%
  \BibitemOpen
  \bibfield  {author} {\bibinfo {author} {\bibfnamefont {A.~P.}\ \bibnamefont {Bartók}}, \bibinfo {author} {\bibfnamefont {R.}~\bibnamefont {Kondor}},\ and\ \bibinfo {author} {\bibfnamefont {G.}~\bibnamefont {Csányi}},\ }\bibfield  {title} {\bibinfo {title} {On representing chemical environments},\ }\href {https://doi.org/10.1103/PhysRevB.87.184115} {\bibfield  {journal} {\bibinfo  {journal} {Physical Review B}\ }\textbf {\bibinfo {volume} {87}},\ \bibinfo {pages} {184115} (\bibinfo {year} {2013})}\BibitemShut {NoStop}%
\bibitem [{\citenamefont {Himanen}\ \emph {et~al.}(2020)\citenamefont {Himanen}, \citenamefont {Jäger}, \citenamefont {Morooka}, \citenamefont {{Federici Canova}}, \citenamefont {Ranawat}, \citenamefont {Gao}, \citenamefont {Rinke},\ and\ \citenamefont {Foster}}]{dscribe}%
  \BibitemOpen
  \bibfield  {author} {\bibinfo {author} {\bibfnamefont {L.}~\bibnamefont {Himanen}}, \bibinfo {author} {\bibfnamefont {M.~O.}\ \bibnamefont {Jäger}}, \bibinfo {author} {\bibfnamefont {E.~V.}\ \bibnamefont {Morooka}}, \bibinfo {author} {\bibfnamefont {F.}~\bibnamefont {{Federici Canova}}}, \bibinfo {author} {\bibfnamefont {Y.~S.}\ \bibnamefont {Ranawat}}, \bibinfo {author} {\bibfnamefont {D.~Z.}\ \bibnamefont {Gao}}, \bibinfo {author} {\bibfnamefont {P.}~\bibnamefont {Rinke}},\ and\ \bibinfo {author} {\bibfnamefont {A.~S.}\ \bibnamefont {Foster}},\ }\bibfield  {title} {\bibinfo {title} {Dscribe: Library of descriptors for machine learning in materials science},\ }\href {https://doi.org/10.1016/j.cpc.2019.106949} {\bibfield  {journal} {\bibinfo  {journal} {Computer Physics Communications}\ }\textbf {\bibinfo {volume} {247}},\ \bibinfo {pages} {106949} (\bibinfo {year} {2020})}\BibitemShut {NoStop}%
\bibitem [{\citenamefont {Horton}\ \emph {et~al.}(2025)\citenamefont {Horton}, \citenamefont {Huck}, \citenamefont {Yang}, \citenamefont {Munro}, \citenamefont {Dwaraknath}, \citenamefont {Ganose}, \citenamefont {Kingsbury}, \citenamefont {Wen}, \citenamefont {Shen}, \citenamefont {Mathis}, \citenamefont {Kaplan}, \citenamefont {Berket}, \citenamefont {Riebesell}, \citenamefont {George}, \citenamefont {Rosen}, \citenamefont {Spotte-Smith}, \citenamefont {{McDermott}}, \citenamefont {Cohen}, \citenamefont {Dunn}, \citenamefont {Kuner}, \citenamefont {Rignanese}, \citenamefont {Petretto}, \citenamefont {Waroquiers}, \citenamefont {Griffin}, \citenamefont {Neaton}, \citenamefont {Chrzan}, \citenamefont {Asta}, \citenamefont {Hautier}, \citenamefont {Cholia}, \citenamefont {Ceder}, \citenamefont {Ong}, \citenamefont {Jain},\ and\ \citenamefont {Persson}}]{mp1}%
  \BibitemOpen
  \bibfield  {author} {\bibinfo {author} {\bibfnamefont {M.~K.}\ \bibnamefont {Horton}}, \bibinfo {author} {\bibfnamefont {P.}~\bibnamefont {Huck}}, \bibinfo {author} {\bibfnamefont {R.~X.}\ \bibnamefont {Yang}}, \bibinfo {author} {\bibfnamefont {J.~M.}\ \bibnamefont {Munro}}, \bibinfo {author} {\bibfnamefont {S.}~\bibnamefont {Dwaraknath}}, \bibinfo {author} {\bibfnamefont {A.~M.}\ \bibnamefont {Ganose}}, \bibinfo {author} {\bibfnamefont {R.~S.}\ \bibnamefont {Kingsbury}}, \bibinfo {author} {\bibfnamefont {M.}~\bibnamefont {Wen}}, \bibinfo {author} {\bibfnamefont {J.~X.}\ \bibnamefont {Shen}}, \bibinfo {author} {\bibfnamefont {T.~S.}\ \bibnamefont {Mathis}}, \bibinfo {author} {\bibfnamefont {A.~D.}\ \bibnamefont {Kaplan}}, \bibinfo {author} {\bibfnamefont {K.}~\bibnamefont {Berket}}, \bibinfo {author} {\bibfnamefont {J.}~\bibnamefont {Riebesell}}, \bibinfo {author} {\bibfnamefont {J.}~\bibnamefont {George}}, \bibinfo {author} {\bibfnamefont {A.~S.}\ \bibnamefont {Rosen}}, \bibinfo {author} {\bibfnamefont
  {E.~W.~C.}\ \bibnamefont {Spotte-Smith}}, \bibinfo {author} {\bibfnamefont {M.~J.}\ \bibnamefont {{McDermott}}}, \bibinfo {author} {\bibfnamefont {O.~A.}\ \bibnamefont {Cohen}}, \bibinfo {author} {\bibfnamefont {A.}~\bibnamefont {Dunn}}, \bibinfo {author} {\bibfnamefont {M.~C.}\ \bibnamefont {Kuner}}, \bibinfo {author} {\bibfnamefont {G.-M.}\ \bibnamefont {Rignanese}}, \bibinfo {author} {\bibfnamefont {G.}~\bibnamefont {Petretto}}, \bibinfo {author} {\bibfnamefont {D.}~\bibnamefont {Waroquiers}}, \bibinfo {author} {\bibfnamefont {S.~M.}\ \bibnamefont {Griffin}}, \bibinfo {author} {\bibfnamefont {J.~B.}\ \bibnamefont {Neaton}}, \bibinfo {author} {\bibfnamefont {D.~C.}\ \bibnamefont {Chrzan}}, \bibinfo {author} {\bibfnamefont {M.}~\bibnamefont {Asta}}, \bibinfo {author} {\bibfnamefont {G.}~\bibnamefont {Hautier}}, \bibinfo {author} {\bibfnamefont {S.}~\bibnamefont {Cholia}}, \bibinfo {author} {\bibfnamefont {G.}~\bibnamefont {Ceder}}, \bibinfo {author} {\bibfnamefont {S.~P.}\ \bibnamefont {Ong}}, \bibinfo
  {author} {\bibfnamefont {A.}~\bibnamefont {Jain}},\ and\ \bibinfo {author} {\bibfnamefont {K.~A.}\ \bibnamefont {Persson}},\ }\bibfield  {title} {\bibinfo {title} {Accelerated data-driven materials science with the materials project},\ }\href {https://doi.org/10.1038/s41563-025-02272-0} {\bibfield  {journal} {\bibinfo  {journal} {Nature Materials}\ }\textbf {\bibinfo {volume} {24}},\ \bibinfo {pages} {1522} (\bibinfo {year} {2025})}\BibitemShut {NoStop}%
\bibitem [{\citenamefont {Jain}\ \emph {et~al.}(2013)\citenamefont {Jain}, \citenamefont {Ong}, \citenamefont {Hautier}, \citenamefont {Chen}, \citenamefont {Richards}, \citenamefont {Dacek}, \citenamefont {Cholia}, \citenamefont {Gunter}, \citenamefont {Skinner}, \citenamefont {Ceder},\ and\ \citenamefont {Persson}}]{mp2}%
  \BibitemOpen
  \bibfield  {author} {\bibinfo {author} {\bibfnamefont {A.}~\bibnamefont {Jain}}, \bibinfo {author} {\bibfnamefont {S.~P.}\ \bibnamefont {Ong}}, \bibinfo {author} {\bibfnamefont {G.}~\bibnamefont {Hautier}}, \bibinfo {author} {\bibfnamefont {W.}~\bibnamefont {Chen}}, \bibinfo {author} {\bibfnamefont {W.~D.}\ \bibnamefont {Richards}}, \bibinfo {author} {\bibfnamefont {S.}~\bibnamefont {Dacek}}, \bibinfo {author} {\bibfnamefont {S.}~\bibnamefont {Cholia}}, \bibinfo {author} {\bibfnamefont {D.}~\bibnamefont {Gunter}}, \bibinfo {author} {\bibfnamefont {D.}~\bibnamefont {Skinner}}, \bibinfo {author} {\bibfnamefont {G.}~\bibnamefont {Ceder}},\ and\ \bibinfo {author} {\bibfnamefont {K.~A.}\ \bibnamefont {Persson}},\ }\bibfield  {title} {\bibinfo {title} {Commentary: The materials project: A materials genome approach to accelerating materials innovation},\ }\href {https://doi.org/10.1063/1.4812323} {\bibfield  {journal} {\bibinfo  {journal} {{APL} Materials}\ }\textbf {\bibinfo {volume} {1}},\ \bibinfo {pages}
  {011002} (\bibinfo {year} {2013})}\BibitemShut {NoStop}%
\bibitem [{\citenamefont {Ong}\ \emph {et~al.}(2008)\citenamefont {Ong}, \citenamefont {Wang}, \citenamefont {Kang},\ and\ \citenamefont {Ceder}}]{mp_phase}%
  \BibitemOpen
  \bibfield  {author} {\bibinfo {author} {\bibfnamefont {S.~P.}\ \bibnamefont {Ong}}, \bibinfo {author} {\bibfnamefont {L.}~\bibnamefont {Wang}}, \bibinfo {author} {\bibfnamefont {B.}~\bibnamefont {Kang}},\ and\ \bibinfo {author} {\bibfnamefont {G.}~\bibnamefont {Ceder}},\ }\bibfield  {title} {\bibinfo {title} {Li-fe-p-o2 phase diagram from first principles calculations},\ }\href {https://doi.org/10.1021/cm702327g} {\bibfield  {journal} {\bibinfo  {journal} {Chemistry of Materials}\ }\textbf {\bibinfo {volume} {20}},\ \bibinfo {pages} {1798} (\bibinfo {year} {2008})}\BibitemShut {NoStop}%
  

\end{thebibliography}
\end{document}